\documentclass[11pt, a4paper,table]{article} 

\usepackage{longtable}
\usepackage{enumitem}
\usepackage{xcolor}
\usepackage{url}
\usepackage[ruled]{algorithm2e}
\providecommand{\keywords}[1]
{
  \small	
  \textbf{\textit{Keywords:}} #1
}

\providecommand{\jel}[1]
{
  \small	
  \textbf{\textit{JEL Classifications:}} #1
}

\SetCommentSty{mycommfont}

\SetKwInput{KwInput}{Input}                
\SetKwInput{KwOutput}{Output}              

\newlist{steps}{enumerate}{1}
\setlist[steps, 1]{resume,leftmargin=*,label = \textbf{Step \arabic*}:}

\newlist{stepss}{enumerate}{1}
\setlist[stepss, 1]{resume,leftmargin=*,label = \textbf{Step \arabic*}:}

\usepackage{lipsum} 
\usepackage[affil-it]{authblk}
\usepackage{enumitem}
\usepackage{blindtext}
\usepackage{multirow}
\usepackage{array}
\usepackage{diagbox}
\usepackage{siunitx}
\newcolumntype{P}[1]{>{\centering\arraybackslash}p{#1}}
\usepackage{caption}
\usepackage[rightcaption]{sidecap} 
\usepackage[para,online,flushleft]{threeparttable}
\usepackage{highlight}
\usepackage{bm}
\renewcommand{\opacity}{80}
\usepackage{hhline} 
\usepackage{subfig}
\usepackage{graphicx}

\usepackage[english]{babel} 
\usepackage{booktabs} 
\usepackage{comment} 
\usepackage[utf8]{inputenc} 
\usepackage[T1]{fontenc} 

\usepackage{amsmath,amsfonts,amsthm} 
\usepackage{tikz} 
\usepackage{tikz-cd}
\usetikzlibrary{positioning,arrows} 
\usetikzlibrary{decorations.pathreplacing} 
\usepackage{tikzsymbols} 
\usepackage{blindtext} 

\usepackage{geometry} 

\usepackage{setspace} 
\usepackage[T1]{fontenc} 
\usepackage[utf8]{inputenc} 

\usepackage{XCharter} 

\usepackage{fancyhdr} 
\usepackage{natbib} 
\setcitestyle{authoryear,open={(},close={)}} 
\usepackage{har2nat} 

\usepackage{rotating}
\usepackage{arydshln}

\title{A Data-driven Explainable Case-based Reasoning Approach for Financial Risk Detection}

\author[1]{Wei Li}
\author[1,2]{Florentina Paraschiv}
\author[3,*]{Georgios Sermpinis}

\affil[1]{\small{NTNU Business School, Norwegian University of Science and Technology, 7491 Trondheim,
Norway}}
\affil[2]{\small{Institute for Operations Research and Computational Finance, University of St. Gallen, Bodanstrasse 6, CH-9000, St. Gallen, Switzerland}}
\affil[3]{\small{Adam Smith Business School, University of Glasgow, G12 8QQ, Glasgow, UK}}

\affil[*]{Correspondence:
Georgios.Sermpinis@glasgow.ac.uk}

\date{}

\begin{document}

\maketitle 
 \renewcommand{\arraystretch}{0.75}

\vspace{-3.5em}
\begin{abstract}
\noindent The rapid development of artificial intelligence methods contributes to their wide applications for forecasting various financial risks in recent years. This study introduces a novel explainable case-based reasoning (CBR) approach without a requirement of rich expertise in financial risk. Compared with other black-box algorithms, the explainable CBR system allows a natural economic interpretation of results. Indeed, the empirical results emphasize the interpretability of the CBR system in predicting financial risk, which is essential for both financial companies and their customers. In addition, our results show that the proposed automatic design CBR system has a good prediction performance compared to other artificial intelligence methods, overcoming the main drawback of a standard CBR system of highly depending on prior domain knowledge about the corresponding field. 
\end{abstract}

 \keywords {Financial risk management, Decision theory, Case-based reasoning, Credit risk, Particle swarm optimization}
 
\jel{C53, C61, C63, D81, G32}    
\setcounter{page}{1} 
\section{Introduction} 
Financial risk is typically associated with the possibility of a loss in the financial field, such as credit risk, operation risk, and business risk. It can have several negative consequences at the firm level, such as the loss of capital to interested stakeholders, and can even affect the economy as a whole, leading to the collapse of the entire financial system. Thus, the financial risk detection (FRD) is vital, and it becomes more important for banks and other financial institutions in the wake of strengthened financial regulations meant to overcome financial crises. Typically, FRD is a classification problem. In recent years, numerous artificial intelligence (AI) classification algorithms have been developed and improved for FRD and achieved considerably accurate results \cite{PENG20112906,Chen2011135, SERMPINIS201819,Hwang2018419,Salim20191569,Eduard2020311,STEVENSON2021758,ZHANG2021}. However, the current generation of AI algorithms has been criticized for being black box oracles that allow limited insight into decision factors. That is, as their mechanism of transforming the input into the output is obfuscated without interference from the users. Thus, black box AI algorithms are not suitable in regulated financial services. Especially, under the rule of the general data protection regulation (GDPR) in Europe, decision-making based solely on automated processing is prohibited, while meaningful information about the logic involved should be carried on \cite{Voigt2017}. To overcome this issue in the financial field, Explainable AI (XAI) models are necessary, which provide reasons to make decisions or enable humans to understand and trust the decisions appropriately.  

CBR is an XAI approach which finds a solution to unravel new problems based on past experiences. In particular, CBR can be formalized as a four-step process \cite{Aamodt1994}: given a new problem (a case without solution), retrieve past solved cases stored in a CBR system similar to the new one; reuse the similar ones to suggest a solution to the new one; revise if the new case is solved; retain the newly solved case in the CBR system. The mechanism of CBR is analogous to a pervasive behavior in human solving problems; whenever encountering a novel problem, humans consider similar situations and adapt a solution from the retrieved case. Thus, it is intuitive that similar cases serve themselves as the explanation in the CBR system to the human users \cite{Sormo2005}. The natural explanation ability of the CBR system boosted its applications in many fields and it is particularly well appreciated by some decision-support systems where there is a preference to understand how the system produces a recommendation \cite{Moxey201025}, such as medical system. \citet{LAMY201942} employ a CBR system for breast cancer diagnosis and explain the therapeutic decision in breast cancer via displaying quantitative and qualitative similarities between the query and similar cases. In the study of \citet{GUESSOUM2014267}, a CBR system is used to promote decision-making of the diagnosis of chronic obstructive pulmonary disease. According to the study of \citet{Brown1994}, CBR performs well in the experience-rich fields, such as diagnosis, prediction, classification, configuration, and planning \citep{ CHI199367, Morris1994, OROARTY1997417, HU201665,MOHAMMED2018212}.

Several prior studies have applied the CBR system in business decision making \cite{Shin1997, Bryant1997, AHN2009599, LI2010137, VUKOVIC20128389, INCE2014205}. In the study of \citet{VUKOVIC20128389}, a CBR method combined with the Genetic algorithm has been used for credit scoring. The experimental results showed that the proposed CBR method improves the performance of the traditional CBR system and outperforms the traditional k-nearest neighbor classifier, but the explainability of CBR method has not been analyzed. \citet{LI2010137} compared the predictive performance of the six hybrid CBR modules in business failure prediction. This study concludes that CBR is preferred over other models because it results in an accurate prediction of a company's financial state. However, the authors did not consider researching on explainability and their research output does not directly propose a strategy to companies that are predicted to fail. In the study of \citet{INCE2014205}, a CBR system was used to select stocks for portfolio optimization, compared with multi layer perceptron, decision trees, generalized rule induction and logistic regression, and showed that the performance of CBR is better than the performance of the other techniques in terms of multiple measures. Similar to other studies, the research neglected the advantages of explanation ability in decision making. In \citet{AHN2009599}, the authors proposed a hybrid CBR model using a genetic algorithm to optimize feature weights to predict bankruptcy, and found that the CBR system has a good explanation ability and high prediction performance over the other AI techniques. However, this study did not conduct an effective empirical analysis to support the arguments.

Prior studies have contributed to the introduction of numerous algorithms for financial risk classification \cite{PENG20112906}. \citet{TSAI20082639} employed the multilayer perceptron (feedforward artificial neural network) for predicting bankruptcy and credit scoring and the empirical results implied that the decision makers should consider the combination of multiple classifiers for bankruptcy prediction and credit scoring rather than a single classifier. \citet{SERMPINIS201819} applied a Lasso regression to predict market implied ratings and found the Lasso models perform better in out-of-sample prediction than ordered probit models. \citet{KAO2012245} proposed a combination of a Bayesian behavior scoring model and a decision tree credit scoring model. The results showed that decision trees can provide critical insights into the decision-making process and that a cardholder's credit history provides significantly important information in credit scoring. The logistic regression and k-nearest neighbor models are traditional classification methods \cite{Henley199677, Bensic2005133}, which are commonly used as benchmark models \cite{WEST20001131, Li2002647, ABDOU20081275, Pavlidis20121645}. Overall, the majority of studies on financial risk prediction focus solely on the accuracy comparison and do not explore the "explainability" feature. 

Successfully developing a CBR system largely depends on an effective retrieval of useful prior cases with the problem. Thus, the integration of domain knowledge and experience about similarity calculation into the case matching and retrieving processes is essential in building a successful CBR model. However, even for experts it is challenging to acquire efficient domain knowledge and define a priori the set of most effective parameters in similarity calculation functions for solving a specific problem. Thus, in the absence of domain knowledge, a data-driven design for the CBR system is in a high demand. Prior research focuses on the optimization of global feature weights \cite{Novakovic2011, Prati2012, Jaiswal2019}. In the study of \citet{Jaiswal2019}, multiple feature scoring methods were discovered to automatically assign the global feature weights of the CBR system in the default detection problem. They showed that the feature scoring data-driven approach was well suited in the initial phases of a CBR system development and provided an opportunity for the developer of the CBR system without domain knowledge. \citet{Novakovic2011} conducted extensive tests on the influence of different feature ranking methods on the performance of classification models. They concluded that the prediction accuracy of the classifiers is determined by the choice of ranking indices. Further, because the characteristics of the input data may differ significantly, no best-ranking index exists for different classifiers under different datasets.

However, in recent research, the knowledge-intensive problem remains in the design process of the CBR system. For instance, the parameters of local similarity functions need knowledge input from the system designers, which requires further research. To fill the research gap, this study aims to develop a data-driven evolutionary CBR system by optimizing local similarity functions with an evolutionary algorithm. In particular, the proposed CBR system is automatically designed without human intervention, yet based on a rigorous selection of inputs. In the experimental study, the designed model is used for FRD and the performance is estimated by employing five categories of financial risk datasets. As indicated in the literature review, the studies \cite{Novakovic2011, Prati2012, Jaiswal2019} that comes closest to ours failed to propose a generalized automatically designed CBR system. Furthermore, prior studies did not check for the CBR predictability in a robustness test, and did not comprehensively analyze and explore the explainability of CBR system in the financial field. Thus, our contributions to the literature are twofold. We first propose a data-driven automatic design CBR system and exhibit its superior performance in FRD. The results show the proposed model performs better than the benchmark AI models, logistic regression, k-nearest neighbor, decision tree, Gaussian Naive Bayes, multi-layer perceptron and lasso regression models. Second, we clarify the four CBR explanation goals, transparency, justification, relevance, and learning, respectively, and display the explainability of the CBR system in a case study of the credit application risk. We are the first to introduce the explanation goals of CBR system in detail with applications for FRD. In addition, we introduce an algorithm for calculating the prediction probability in the CBR system to justify the prediction results.  

The rest of this paper is organized as follows. Section 2 introduces the proposed evolutionary CBR system and presents the explanation goals of CBR. Section 3 describes the detailed experiment of FRD. The experiment results are shown in Section 4. Section 5 concludes the paper.

\section{Methodology}
\subsection{Evolutionary CBR}
\subsubsection{The local-global principle for similarity measures}
The CBR system is designed to find the most similar cases of a query case in the database. In the process of retrieving, similarity measures play a vital role in assigning a degree of similarity to cases. Typically, the local-global principle is widely used in the attribute-based CBR system for case representation and similarity calculation \cite{Michael2013}. In general, the global similarity is measured by the square root of the weighted sum of all the local similarities. Given a query case $Q$ and a case $C$ from $L$-dimensional database ($L$ features), a global similarity function $Sim(Q, C)$ to calculate the similarity between $Q$ and $C$ can be described as follows:
\begin{equation}
    Sim(Q,C) = \sqrt{\sum_{j=1}^L \bm{w_j}\times(sim_j(q_j,c_j))^2}
\end{equation}
\noindent where, for the attribute $j$, $sim_j$ is the local similarity function, $q_j$ and $c_j$ are attribute value from the case Q and C, respectively. $\bm{w_j}$ stands for the weight (global parameters) of the attribute $j$. 

\noindent For the local (feature) similarity, asymmetrical polynomial functions are commonly used to measure the similarity of attribute-value \cite{Bach201217}. It can be represented as:

  \begin{equation}
   sim_j(q_j,c_j)=
    \begin{cases}
      \big(\frac{D_{j}-(c_j-q_j)}{D_{j}}\big)^{\bm{a_j}}, & \text{if}\ q_j \leq c_j \\
      \big(\frac{D_{j}-(q_j-c_j)}{D_{j}}\big)^{\bm{b_j}}, & \text{if}\ q_j > c_j
    \end{cases}
  \end{equation}
  \noindent where $D_{j}$ stands for the difference between maximum and minimum value of attribute $j$ in dataset. $\bm{a_j}$ and $\bm{b_j}$ are the degree (local parameters) of polynomial functions. A simple instance of the similarity calculation can be found in Appendix \ref{Instance}.
  
\subsubsection{Data-driven automatic CBR design}
In the proposed evolutionary CBR framework (CBR\_E), classification works by calculating the similarities between a query case and all the cases in a dataset based on the equation 1 and 2, selecting a specified amount (\bm{$k$}) of cases most similar to the query case. Then, a majority voting is used to assign the query case the most common class among its \bm{$k$} most similar cases. Thus, the parameter \bm{$k$} is the other parameter required, associated with the global parameter $\bm{w_j}$ and local parameters $\bm{a_j}$ and $\bm{b_j}$, for automatically designing a data-driven CBR system without human involvement. 

For obtaining the parameter \bm{$k$}, the well-known k nearest neighbors algorithm (KNN) is employed, which can be considered as a non-parametric CBR. Typically, a case is classified by a plurality vote of its \bm{$k$} distance-based neighbors in the KNN paradigm. Indeed, the \bm{$k$} is the only parameter influencing the classification accuracy of KNN model, which is required to be determined. For a specific dataset, the optimal \bm{$k$} can be obtained by cross-validated grid search over a parameter grid.

The weights $\bm{w_j}$ reflect the influence of the attributes on the global measure. In order to calculate the importance of the attributes, the feature importance scoring methods are employed. The scores of attributes will be transformed into the global weights, $\bm{w_j}$, in the CBR system by scaling to sum to 1. In this study, six scoring methods are applied to generate six sets of global weights $\bm{w_j}$, which are Gini \cite{Ceriani2012}, Information entropy \cite{Kullback59}, Mutual information \cite{Kraskov2004}, Chi2 \cite{Cost1993}, ANOVA \cite{LIN201164} and ReliefF \cite{Kononenko1997}. Consequently, we create six CBR systems based on the generated weights. For each created CBR system, the optimal local parameters, $\bm{a_j}$ and $\bm{b_j}$ of polynomial functions, are searched by particle swarm optimization (PSO) algorithm, in which the cost function is the classification accuracy. The explanation of PSO can be found in Appendix \ref{PSO}. 

After evaluating the performance of the six designed CBR systems through cross-validation\footnote{The computation is based on parallel computing, explained in Appendix \ref{Parallel_computing}}, the best-validated one will be selected and used for financial risk prediction. The designing process of the proposed CBR system can be described as follows:\\

\begin{algorithm}[H]
  
\KwInput{Financial data input}
\KwOutput{Designed CBR system}
Data processing. \\
Determine the number of the most similar cases \bm{$k$} for retrieval with KNN algorithm.\\
\While{There are more feature scoring methods}
  {
    Score the features and assign the weights $\bm{w_j}$ using the feature importance scoring method. \\
    Optimize the parameters, $\bm{a_j}$ and $\bm{b_j}$, of the local similarity functions using PSO algorithm. \\
    Evaluate the CBR system via cross-validation.\\ 
    }
Compared the performance of all the trained CBR systems, select the best-validated one. \\
\caption{Data-driven CBR system design}
\end{algorithm}
\vspace{2mm}
\subsection{Explainability}
Explanations differ in terms of explanation goals. In the CBR system, four major goals of explanation are provided \cite{Sormo2004165}: transparency, justification, relevance, and learning.   

\subsubsection{Explain how the system reached the answer (Transparency)}
The goal of the explanation of transparency is to allow users to understand and examine how the system finds an answer. It is fairly intuitive to understand the basic concept of retrieving similar and concrete cases to solve the current problem. This understanding supports the basic approach in CBR explanation, which is to display the most similar cases to the present case, compare them, explain the decision-making process, and explore the reasons of the default \cite{Sormo2005}. In addition, some research has shown that the explanation of predictions is important, and case-based explanations will significantly improve user confidence in the solution compared to the rule-based explanations or only displaying the problem solution \cite{Cunningham2003122}.

\subsubsection{Explain why the answer is a good answer (Justification)} \label{Probability}
The justification goal is to increase the confidence in the solution provided by the system by offering some supports. For instance, the posterior probability is usually important in the classification problem, which gives a confidence measure in the classification result. Similar to KNN \cite{Atiya2005731}, the CBR system can provide a posterior probability estimator. In our case, financial risk detection is a binary classification problem with classes $Y$ ($Y=0$ (non-default) or $Y=1$ (default)). Assume a dataset $X$ includes $N$ labeled cases $x(n)$, $n = 1, ..., N$, and for a query case $x$, $K'$ is the number in the $K$ most similar cases belong to the class default ($Y = 1$), the estimate of the default probability $\hat{P}(Y=1|X=x)$ is given by 
\begin{equation}
\hat{P}(Y = 1|X = x) = \frac{K'}{K}
\end{equation}
However, it is not intuitive to consider that every case in the $k$ most similar cases has the same weights. The more similar case should have a higher contribution to the probability calculation than the less similar case. Thus, it is better to generalize this estimator by assigning different probabilities to the different similar cases. Let the probabilities assigned to the $k$ most similar cases be $p_1$, ..., $p_K$ and the label $B_i=1$ if the $i^{th}$ case belongs to the class $Y=1$ and $B_i=0$ otherwise. These probabilities are greater than or equal zero, monotonically decreasing, and sum to 1: $\sum_{i=1}^kp_i=1$ (constraints). Then the probability estimate of the default is given: 
\begin{equation}
\hat{P}(Y = 1|X = x) = \sum_{i=1}^{K+1}  B_i \times p_i
\end{equation}
The optimal probabilities $p_1$, ..., $p_{K+1}$ are determined by maximizing the likelihood of the dataset $X$. It is worth to note that the $K+1$ probabilities rather than $K$ probabilities are used. The $B_{K+1} \times p_{K+1}$ is a regularization term to prevent obtaining $-\infty$ log likelihood by assigning $B_{K+1}=1/2$. Further, to reduce the constraints when optimizing log likelihood function to obtain the probabilities, a softmax representation is used:  
\begin{equation}
p_i = \frac{e^{\omega_i}}{\sum_{j=1}^{K+1}e^{\omega_j}},\;\; \text{ for } i = 1 ,..., K+1
\end{equation}
where the parameters $\omega_i$ can be any value and constrained by monotonically decreasing. Then, the estimate function of the default probability becomes:  
\begin{equation}
\hat{P}(Y=1|X = x) = \frac{\sum_{i=1}^{K+1}B_{i}e^{\omega_i}}{\sum_{j=1}^{K+1}e^{\omega_j}}
\end{equation}
Let $B(n)$ denotes the class membership of $x(n)$. The likelihood $\mathcal{L}$ of the $N$ cases dataset $X$ is: 
\begin{equation}
\mathcal{L}= \prod_{n=1}^N \hat{P}(Y=1|X=x(n)) =  \prod_{n=1}^N \left[\frac{\sum_{i=1}^{K+1}B_i(n)e^{\omega_i}}{\sum_{j=1}^{K+1}e^{\omega_j}}\right]
\end{equation}
where the different probability estimates are assumed to be independent as the dependent case is complicated to analyze. Finally, the log likelihood is given:
\begin{equation}
\log(\mathcal{L}) = \sum_{n=1}^N\log\left[\frac{\sum_{i=1}^{K+1}B_i(n)e^{\omega_i}}{\sum_{j=1}^{K+1}e^{\omega_j}}\right]
\end{equation}
subject to the constraint:
\begin{equation*}
  \omega_1 \geq \omega_2 \geq \omega_3\geq ... \geq \omega_k
\end{equation*}
The optimal weighting parameters $\omega_j$ are determined by maximizing the log likelihood function.

\subsubsection{Which information was relevant for the decision making process (Relevance)}
Different information input has different contributions to solve the problem. Identification of most relevant information can be used to adjust the options of financial companies regarding the future direction of a business operation. CBR system allows users to recognize which factors are important for decision making by analyzing the global weights. 

In the proposed evolutionary CBR system, the weights are automatically calculated by applying feature scoring methods. For instance, the Gini index is used to rank the features and determine which features are the most relevant information in a dataset. In addition, the Gini index is commonly used to split a decision tree, such as C4.5 \cite{Quinlan1993}, which can be combined with the CBR system to diagnose the reasons for the problem. In our study, we consider a technique, Cause Induction in Discrimination Tree (CID Tree) \cite{Radhika2005786}, to identify the possible features that could be causally linked to the default case. In particular, the algorithm aims to select pairs of nodes, P and S, which have high importance with respect to discriminating alternative classes (default and non-default). The relevance score of each node for causing the default is given by the following \cite{Radhika2005786}: 
\begin{equation}\label{cid_eq}
  \mathrm{Score_p} = (\mathrm{D_p^2 \times N_s^2)/(N_p^2 \times D_s^2)}
\end{equation}
where N\textsubscript{p}, D\textsubscript{p} stand for the number of good and bad cases under the same parent node P. N\textsubscript{s}, D\textsubscript{s} stand for the number of good and bad cases under the sibling node S. The relevance score only depends on the number of cases under the node P and S. The score of the node P is high when it has a relatively high number of default cases compared to the node S. The higher the score is, the more likely the default can be discriminated.

\subsubsection{Which information can be explored based on the current situation (Learning)}
This goal aims to not merely find a good solution to a problem and explain the solution to the financial companies but explore new information and deepen their understanding of the domain knowledge. The information can guild financial companies better analyze and solve the problems. 

Integration of data mining techniques with prediction methods can lead to better analysis of the domain knowledge and extracting useful relationships in data to improve the decision making process \cite{Aamodt98combiningcase, Arshadi20051127, Gouttaya2012136}. Data mining techniques typically involve a process of exploring and analyzing data and transform it to useful information, which can be used in a variety of tasks \cite{Fayyad199637}. Among these techniques, clustering is a commonly applied method to discover groups and structures in the data. Typically, in the field of market research, clustering is an effective and frequently used method for market segmentation as the same segmented groups of customers tend to have certain similarities and common characteristics \cite{10.1145/1089551.1089610}. Based on the research on customer segmentation, companies can find out targeted market and groups of customers effectively and appropriately. In our study, we apply k-means \cite{LIKAS2003451} as the clustering algorithm to detect more useful information for decision making.

\section{Experiment}
This study employs a multiple-criteria decision-making (MCDM) method to rank the selected classification models based on experimental results. In this section, the experimental study is described in four aspects: benchmark models, data description, performance measure, and experiment design.
\subsection{Benchmark models}
As aforementioned, financial risk prediction is a classification problem and has been explored in several prior studies such as \cite{WEST20001131, Li2002647, Bensic2005133, TSAI20082639, PENG20112906, KAO2012245, SERMPINIS201819}. In this experiment, six well-known classifiers are used as benchmark models, namely logistic regression (LR), k-nearest neighbor (KNN), decision tree (DT), Gaussian Naive Bayes (GNB), multi-layer perceptron (MLP) and lasso regression (LASSO). The naive benchmark is an equally-weighted CBR (CBR\_EW). The introduction of the benchmark models can be found in Appendix \ref{Benchmark_models}. In particular, the features are globally treated with equal importance (\bm{$w_j$} = 1/\bm{$k$}) and locally linear related (\bm{$a_j$} = 1 and \bm{$b_j$} = 1) when constructing the CBR model.  

\subsection{Data}
The characteristics of datasets, such as size and class distribution, can affect the performance of models. Thus, we consider five different financial risk datasets to evaluate the performance of the classification algorithms. The datasets applied in this experiment are collected from the databases UCI and Kaggle, presenting five aspects of financial risk: credit card fraud (CCF), credit card default (CCD), south German credit (SGC), bank churn (BC), and financial distress (FD). The datasets are imbalanced and their statistics are shown in Table \ref{t:datasets}.

\begin{table}[ht!]
\caption{Statistics of the datasets used in the experiment.}\label{t:datasets}
 \begin{threeparttable}
 \begin{tabular*}{\textwidth}{l l P{1.5cm} P{2cm} P{2cm} P{1.5cm} P{1.5cm}  } 
\toprule
Dataset&Acronym&Instances&Positive&Negative&Features&Source\\
\midrule
Credit card fraud&CCF&284,807&492&284,315&30&Kaggle\\
Credit card default&CCD&30,000&6,636&23,364&24&UCI \\
South German credit &SGC&1,000&700&300&20& UCI \\
Bank churn&BC&10,127&1,627&8,500&10&Kaggle\\
Financial distress&FD&3,672&136&3,536&84&Kaggle\\
\bottomrule
\end{tabular*}

 \begin{tablenotes}
 \scriptsize
      \item Notes: Positive indicates an instance is detected as an abnormal case, like the default of credit card bill payment. Negative stands for that an instance is detected as a normal case.  
    \end{tablenotes}
  \end{threeparttable}
\end{table}

\subsubsection{Data description}
\noindent\underline{Credit card fraud dataset \cite{DALPOZZOLO20144915}}\\
The credit card fraud dataset contains transactions made by credit cards by European cardholders. It is important to recognize fraudulent transactions for credit card firms to protect their customers not to be charged for items that they did not purchase. Due to confidentiality issues, the original features and more background information about the data are not provided.\\

\noindent\underline{Credit card default dataset \cite{YEH20092473}}\\
The credit card default dataset was collected from credit card clients in Taiwan. Credit card default happens when the cardholders have become severely delinquent on the credit card payments. Default is a serious credit card status, leading to the loss of creditor and harming credit card customer's ability to get approved for other credit-based services. The predictor variables contain information on default payments, demographic factors, credit data, history of payment, and bill statements.\\

\noindent\underline{South German credit dataset \cite{Ulrike2019}}\\
In the south German credit dataset, each entry represents a person who takes credit from a German bank. The original dataset includes twenty categorial/symbolic attributes. The predictor attributes describe the status of an existing checking account, credit history, duration, education level, employment status, personal status, age, and so on. \\

\noindent\underline{Bank churn dataset \cite{Rahman2020}}\\
The bank churn dataset is applied to predict which customers will leave a bank. The analysis of bank churn is advantageous for banks to recognize what leads a client towards the decision to churn. The attributes of the dataset contain credit score, customers' tenure, age, gender, and so on.\\

\noindent\underline{Financial distress dataset \cite{Ebrahimi2017}}\\
The financial distress dataset is used to make a prediction for the financial distress of a sample of companies. Financial distress is a situation when a corporate cannot generate sufficient revenues, making it unable to cover its financial obligation. In the dataset, the features are some financial and non-financial characteristics of the sampled companies. The names of the features in the dataset are confidential. 

\subsubsection{Data balancing }
The random under-sampling method is applied in this study. This method involves randomly selecting cases from the majority class and remove them from the training dataset until a balanced distribution of classes is reached. Ten random balanced samples are prepared for the performance evaluation of classification models.

\subsection{Performance measure}
The evaluation of learned models is one of the most important problems in financial risk detection. Typically, the performance metrics used in evaluating classification models include: (1) Overall accuracy (2) Precision (3) Recall (4) Specificity (5) F1-score (6) ROC\_AUC (7) G-mean. For instance, overall accuracy is the percentage of correctly classified individuals. It is the most common and simplest measure to evaluate a classifier.
\begin{equation}
    Accuracy = \frac{TP + TN}{TP+FN+FP+TN}
\end{equation}
where TP (true positive) is the number of correctly classified positive instances. TN (true negative) is the number of correctly classified negative instances. FP (false positive) is the number of positive instances misclassified. FN (false negative) is the number of negative instances misclassified. The description of the rest of measure metrics can be found in Appendix \ref{Measure_metrics}. Those measures have been developed for various evaluating targets and can show different evaluation results for classifiers given a dataset. Thus, a comprehensive performance metric is required to be applied to evaluate the quality of models. 
\subsubsection{Technique for order preference by similarity to ideal solution (TOPSIS)}
MCDM method is used to evaluate classification algorithms over multiple criteria \cite{Brunette2009}. TOPSIS, a widely used MCDM, is conducted in the experiment. The procedure of TOPSIS can be summarized in Appendix \ref{TOPSIS}. 

The paired $t$ tests are conducted to obtain the performance scores used in TOPSIS. In particular, it compares the classification performance of ten random balanced samples for an individual measure of two classifiers. If their performance is different at the statistically significant 5 \% level, the performance score of the better model is assigned to 1, and the other is -1. Otherwise, both their performance scores are 0. The comparison process is conducted for each measure in each dataset. The sum of performance scores from all datasets is the performance score of a classifier for a given measure metric. Similar MCMD evaluation procedures were conducted in the literature \cite{PENG20112906, Song201984897}.

\subsection{Experimental design}
The dataset is apportioned into train and test sets, with an 80-20 split. The 5-fold cross-validated grid-search is used to optimize the models. Based on the introduction above, the process of evaluating the classification models can be described as follows:
\begin{stepss}
 \item Remove the input data with missing values and normalize the data to the range [0,1].
 \item Apply the random under-sampling method to generate ten balanced samples for each financial dataset.
 \item Train and test multiple classification models and get the measure performances for each generated sample. 
 \item Calculate the performance scores with paired $t$ tests.
 \item Conduct TOPSIS method to evaluate the relative performance of the classification models.

\end{stepss} 

\section{Results}
\subsection{Empirical evidence}
The classification results of eight classifiers on the five financial datasets, evaluated by seven measure metrics, are reported in Table \ref{t:scores_performance}. The results are calculated in terms of the average measure performance of ten randomly balanced samples of each dataset. The best result of a specific measure in a specific dataset is highlighted in boldface, and the performance is column-wise colored (the redder, the better). From Table \ref{t:scores_performance}, no classifier performs the best across all measures for a single dataset or has the best performance for a single measure across all datasets. The results are aligned with the observations from the study of \citet{Novakovic2011}. However, we can observe that CBR method clearly shows competitive performance among the measures and performs stable across the different data sets. As there is no obvious large gradient variation in most colored columns of datasets, the detection of the statistically significant differences between the performance of two classifiers by the $t$ test is important. 
\begin{table*}[ht!]
\captionsetup{font=scriptsize}
\caption{Classification results.}\label{t:scores_performance}
 \begin{threeparttable}
 \begin{tabular*}{\textwidth}{>{\scriptsize}P{1.cm} >{\scriptsize}{l} >{\scriptsize}P{1.4cm} >{\scriptsize}P{1.4cm}>{\scriptsize} P{1.4cm}>{\scriptsize} P{1.4cm} >{\scriptsize}P{1.5cm}>{\scriptsize} P{1.5cm}>{\scriptsize} P{1.5cm} } 
\toprule 
\multirow{2}{*}{\scriptsize{Dataset}}&\multirow{2}{*}{\scriptsize{Algorithm}}&\multicolumn{7}{c}{\scriptsize{Measure}}\\
\cmidrule{3-9}
&&Accuracy&Precision&Recall&Specificity&F1-score&ROC\_AUC&G-mean\\
\midrule
\multirow{8}{*}{\scriptsize{CCF}}&LR&\gradientccf{0.9401}&\gradientccf{0.9739}&\gradientccfl{0.9173}&\gradientccf{0.9690}&\gradientccf{0.9448}&\gradientccf{0.9431}&\gradientccf{0.9428}\\
&KNN     &\gradientccf{0.9102}&\gradientccf{0.9812}&\gradientccf{0.8555}&\gradientccf{0.9793}&\gradientccf{0.9140}&\gradientccf{0.9174}&\gradientccf{0.9153}\\
&DT      &\gradientccf{0.9198}&\gradientccf{0.9511}&\gradientccf{0.9027}&\gradientccf{0.9414}&\gradientccf{0.9263}&\gradientccf{0.9221}&\gradientccf{0.9219}\\
&GNB     &\gradientccf{0.9056}&\gradientccf{0.9711}&\gradientccf{0.8564}&\gradientccf{0.9678}&\gradientccf{0.9101}&\gradientccf{0.9121}&\gradientccf{0.9104}\\
&MLP     &\gradientccf{0.8949}&\gradientccf{0.9715}&\gradientccf{0.8364}&\gradientccf{0.9690}&\gradientccf{0.8989}&\gradientccf{0.9027}&\gradientccf{0.9002}\\
&LASSO     &\gradientccf{0.8970}&\gradientccfl{0.9880}&\gradientccf{0.8255}&\gradientccfl{0.9874}&\gradientccf{0.8995}&\gradientccf{0.9064}&\gradientccf{0.9028}\\
&CBR\_EW &\gradientccf{0.9213}&\gradientccf{0.9807}&\gradientccf{0.8764}&\gradientccf{0.9782}&\gradientccf{0.9256}&\gradientccf{0.9273}&\gradientccf{0.9259}\\
&CBR\_E  & \gradientccfl{0.9406}&\gradientccf{0.9776}&\gradientccf{0.9145}&\gradientccf{0.9736}&\gradientccfl{0.9450}&\gradientccfl{0.9441}&\gradientccfl{0.9436}\\[0.1cm]
\cmidrule{2-9}
\multirow{8}{*}{\scriptsize{CCD}}&LR&\gradientccd{0.6715}&\gradientccd{0.6778}&\gradientccd{0.6368}&\gradientccd{0.7052}&\gradientccd{0.6567}&\gradientccd{0.6710}&\gradientccd{0.6701}\\
&KNN     &\gradientccd{0.6610}&\gradientccd{0.6914}&\gradientccd{0.5652}&\gradientccd{0.7543}&\gradientccd{0.6220}&\gradientccd{0.6598}&\gradientccd{0.6530}\\
&DT      &\gradientccd{0.6968}&\gradientccd{0.7428}&\gradientccd{0.5898}&\gradientccd{0.8011}&\gradientccd{0.6575}&\gradientccd{0.6954}&\gradientccd{0.6874}\\
&GNB     &\gradientccd{0.6205}&\gradientccd{0.5823}&\gradientccdl{0.8166}&\gradientccd{0.4294}&\gradientccdl{0.6798}&\gradientccd{0.6230}&\gradientccd{0.5921}\\
&MLP     &\gradientccdl{0.6994}&\gradientccdl{0.7464}&\gradientccd{0.5917}&\gradientccdl{0.8042}&\gradientccd{0.6601}&\gradientccdl{0.6980}&\gradientccdl{0.6898}\\
&LASSO    &\gradientccd{0.6730}&\gradientccd{0.6838}&\gradientccd{0.6273}&\gradientccd{0.7175}&\gradientccd{0.6544}&\gradientccd{0.6724}&\gradientccd{0.6709}\\
&CBR\_EW &\gradientccd{0.6658}&\gradientccd{0.6760}&\gradientccd{0.6198}&\gradientccd{0.7106}&\gradientccd{0.6466}&\gradientccd{0.6652}&\gradientccd{0.6636}\\
&CBR\_E  &\gradientccd{0.6844}&\gradientccd{0.6978}&\gradientccd{0.6356}&\gradientccd{0.7318}&\gradientccd{0.6653}&\gradientccd{0.6837}&\gradientccd{0.6820}\\[0.1cm]
\cmidrule{2-9}
\multirow{8}{*}{\scriptsize{SGC}}&LR&\gradientgca{0.7042}&\gradientgca{0.6975}&\gradientgcal{0.7034}&\gradientgca{0.7049}&\gradientgcal{0.7004}&\gradientgca{0.7042}&\gradientgca{0.7042}\\
&KNN     &\gradientgca{0.6592}&\gradientgca{0.6698}&\gradientgca{0.6051}&\gradientgca{0.7115}&\gradientgca{0.6358}&\gradientgca{0.6583}&\gradientgca{0.6561}\\
&DT      &\gradientgca{0.6733}&\gradientgcal{0.7260}&\gradientgca{0.5390}&\gradientgcal{0.8033}&\gradientgca{0.6187}&\gradientgca{0.6711}&\gradientgca{0.6580}\\
&GNB     &\gradientgca{0.7017}&\gradientgca{0.7086}&\gradientgca{0.6678}&\gradientgca{0.7344}&\gradientgca{0.6876}&\gradientgca{0.7011}&\gradientgca{0.7003}\\
&MLP     &\gradientgca{0.6725}&\gradientgca{0.6684}&\gradientgca{0.6627}&\gradientgca{0.6820}&\gradientgca{0.6655}&\gradientgca{0.6723}&\gradientgca{0.6723}\\
&LASSO     &\gradientgcal{0.7075}&\gradientgca{0.7071}&\gradientgca{0.6915}&\gradientgca{0.7230}&\gradientgca{0.6992}&\gradientgcal{0.7072}&\gradientgcal{0.7071}\\
&CBR\_EW &\gradientgca{0.6575}&\gradientgca{0.6479}&\gradientgca{0.6644}&\gradientgca{0.6508}&\gradientgca{0.6561}&\gradientgca{0.6576}&\gradientgca{0.6576}\\
&CBR\_E  &\gradientgca{0.6658}&\gradientgca{0.6562}&\gradientgca{0.6729}&\gradientgca{0.6590}&\gradientgca{0.6644}&\gradientgca{0.6659}&\gradientgca{0.6659}\\[0.1cm]
\cmidrule{2-9}
\multirow{8}{*}{\scriptsize{BC}}&LR&\gradientbc{0.6907}&\gradientbc{0.6988}&\gradientbc{0.6743}&\gradientbc{0.7071}&\gradientbc{0.6863}&\gradientbc{0.6907}&\gradientbc{0.6905}\\
&KNN     &\gradientbc{0.6928}&\gradientbc{0.7155}&\gradientbc{0.6438}&\gradientbc{0.7421}&\gradientbc{0.6777}&\gradientbc{0.6929}&\gradientbc{0.6912}\\
&DT      &\gradientbc{0.7210}&\gradientbc{0.7481}&\gradientbc{0.6694}&\gradientbc{0.7729}&\gradientbc{0.7066}&\gradientbc{0.7212}&\gradientbc{0.7193}\\
&GNB     &\gradientbc{0.7244}&\gradientbc{0.7415}&\gradientbc{0.6922}&\gradientbc{0.7569}&\gradientbc{0.7160}&\gradientbc{0.7245}&\gradientbc{0.7238}\\
&MLP     &\gradientbc{0.7491}&\gradientbc{0.7372}&\gradientbcl{0.7770}&\gradientbc{0.7209}&\gradientbcl{0.7566}&\gradientbc{0.7490}&\gradientbc{0.7485}\\
&LASSO   &\gradientbc{0.6893}&\gradientbc{0.6983}&\gradientbc{0.6707}&\gradientbc{0.7081}&\gradientbc{0.6842}&\gradientbc{0.6894}&\gradientbc{0.6891}\\
&CBR\_EW &\gradientbc{0.7134}&\gradientbc{0.7339}&\gradientbc{0.6729}&\gradientbc{0.7542}&\gradientbc{0.7020}&\gradientbc{0.7135}&\gradientbc{0.7124}\\
&CBR\_E  &\gradientbcl{0.7539}&\gradientbcl{0.7668}&\gradientbc{0.7323}&\gradientbcl{0.7756}&\gradientbc{0.7491}&\gradientbcl{0.7539}&\gradientbcl{0.7536}\\[0.1cm]
\cmidrule{2-9}
\multirow{8}{*}{\scriptsize{FD}}&LR&\gradientfd{0.8273}&\gradientfd{0.7798}&\gradientfd{0.8640}&\gradientfd{0.7967}&\gradientfd{0.8197}&\gradientfd{0.8303}&\gradientfd{0.8297}\\
&KNN     &\gradientfd{0.7382}&\gradientfd{0.7054}&\gradientfd{0.7280}&\gradientfd{0.7467}&\gradientfd{0.7165}&\gradientfd{0.7373}&\gradientfd{0.7373}\\
&DT      &\gradientfd{0.8255}&\gradientfd{0.7619}&\gradientfd{0.8960}&\gradientfd{0.7667}&\gradientfd{0.8235}&\gradientfd{0.8313}&\gradientfd{0.8288}\\
&GNB     &\gradientfd{0.6273}&\gradientfd{0.6800}&\gradientfd{0.3400}&\gradientfdl{0.8667}&\gradientfd{0.4533}&\gradientfd{0.6033}&\gradientfd{0.5428}\\
&MLP     &\gradientfd{0.8200}&\gradientfdl{0.8186}&\gradientfd{0.7760}&\gradientfd{0.8567}&\gradientfd{0.7967}&\gradientfd{0.8163}&\gradientfd{0.8153}\\
&LASSO     &\gradientfd{0.8327}&\gradientfd{0.7705}&\gradientfdl{0.9000}&\gradientfd{0.7767}&\gradientfdl{0.8303}&\gradientfd{0.8380}&\gradientfd{0.8361}\\
&CBR\_EW &\gradientfd{0.7582}&\gradientfd{0.7175}&\gradientfd{0.7720}&\gradientfd{0.7467}&\gradientfd{0.7437}&\gradientfd{0.7593}&\gradientfd{0.7592}\\
&CBR\_E  &\gradientfdl{0.8364}&\gradientfd{0.7963}&\gradientfd{0.8600}&\gradientfd{0.8167}&\gradientfd{0.8269}&\gradientfdl{0.8383}&\gradientfdl{0.8381}\\[0.1cm]
\bottomrule
\end{tabular*}
 \begin{tablenotes}
      \scriptsize
      \item  Notes: Color indicates the performance
columnwise (the redder, the better). For all statistical measures retained the higher the value, the more is the corresponding model. The best values of each column are depicted in bold.
    \end{tablenotes}
  \end{threeparttable}
\end{table*}

The performance scores of all classification models are calculated, based on the measure results in Table \ref{t:scores_performance}, are shown in Table \ref{t:ttest}. For each performance measure, the best score is highlighted in boldface. The higher performance score indicates the classifier performs statistically significantly better than the others for a specific measure over five financial datasets. However, no classifier has the best performance for all measures. The results are consistent with the ones also reported in the research of  \citet{PENG20112906}. Therefore, the MCDM method is required to provide an overall ranking of classification algorithms. 

\begin{table*}[ht!]
\caption{Performance scores of algorithms.}\label{t:ttest}
 \begin{threeparttable}
 \begin{tabular*}{\textwidth}{l P{1.4cm} P{1.4cm} P{1.4cm} P{1.6 cm} P{1.45cm} P{1.5cm} P{1.5cm} } 
\toprule
\diagbox{\scriptsize{Algorithm}}{\scriptsize{Measure}}&Accuracy&Precision&Recall&Specificity&F1-score&ROC\_AUC&G-mean\\
\midrule
LR& 7&  -3&  12& -10&   6&  7&8\\
KNN&-18&  -8& -19&   3& -21& -18& -19\\
DT& 7&   8&  -5&   \textbf{11}&   0&   6&   4\\
GNB&-13& -4&  -3&  -5&  2& -13& -11\\
MLP& 13&   \textbf{10}&  1&   7&  6&   13&   10\\
LASSO&-3&  -1&  2&   -3&  -3& -2& -1\\
CBR\_EW& -9& -8& -5&  -5& -8& -9& -9\\
CBR\_E&\textbf{16}&   6&  \textbf{17}&   2&  \textbf{18}&  \textbf{16}&  \textbf{18}\\
\bottomrule
\end{tabular*}
 \begin{tablenotes}
      \scriptsize
      \item Notes: With bold, the best value is depicted in each column. The higher performance score indicates the classifier performs statistically significantly better than the others for a specific measure over five financial datasets.
    \end{tablenotes}
  \end{threeparttable}
\end{table*}
The ranking of the classification models generated by TOPSIS is shown in Table \ref{t:topsis}. From the table, we can see the proposed CBR has the relative best performance. Compared with the naive benchmark equally-weighted CBR model, the performance of the proposed CBR model has been improved considerably. In summary, we can conclude that the proposed data-driven evolutionary CBR has an overall better performance than the other AI classification algorithms for financial risk prediction problems.

\begin{table}
\caption{TOPSIS values. }\label{t:topsis}
 \begin{threeparttable}
 \begin{tabular*}{\textwidth}{p{2.5cm} P{11cm} P{1.5cm}} 
\toprule
Algorithm&TOPSIS&Ranking\\
\midrule
CBR\_E& 0.8379&1\\
MLP& 0.7561&2\\
DT&0.6475&3\\
LR&0.5685 &4\\
LASSO&0.4593&5\\
GNB&0.3342&6\\
CBR\_EW& 0.2805&7\\
KNN& 0.2024&8 \\
\bottomrule
\end{tabular*}
 \begin{tablenotes}
      \scriptsize
      \item Notes: The table presents the TOPSIS values of all models under study and their related ranking. Higher TOPSIS value is associated with better model performance.
    \end{tablenotes}
  \end{threeparttable}
\end{table}

\subsection{Interpretation of results}
Compared with the other classification methods, one of the important characteristics of CBR is the interpretability of the prediction result. In this section, we conduct a case study with the dataset of the south German credit to show the explainability of the CBR system. The south German credit dataset is publicly available and widely used in the scientific field for research on credit risk prediction, such as the recent studies of \citet{Ha2019511}, \citet{Alam2020201173} and \citet{TRIVEDI2020101413}. The dataset provider offered a detailed description of the features, which are essential information to explain the results. In contrast, the other public datasets used in this paper are either names of features are confidential, or a description of features is missing, which makes them not suitable for explainability study. Thus, we use the German credit dataset to perform the case study. In this dataset, each entry represents a person who takes credit from a bank. Each person is classified as subject to credit risk or not according to the set of features. The detailed description of features can be found in Appendix \ref{german_data}.

\subsubsection{Results explanation based on similar cases}
Applying the CBR system, a bank can provide reasons/suggestions for consumers who failed in applying for credit from the bank. As aforementioned, case-based explanation will promote confidence in the decision. 

For instance, an application case $C_0$ has been correctly classified as a bad credit risk by applying the CBR system (voting from the three most similar cases, the optimal \bm{$k$} = 3). Through similarity queries to the cases' base, the three most similar cases of the case $C_0$, $C_1^d$ (default), $C_2^d$ (default) and $C_3^n$ (non-default), and its second most similar good credit risk cases $C_4^n$ (non-default) can be found and their attributes are shown in Table \ref{t:feature_example}. From Table \ref{t:feature_example}, we can observe that the difference between $C_0$ and its most similar case $C_1^d$ is that the latter has less credit amount and duration. The repayment default of the case $C_1^d$ has a strong indication that case $C_0$ will default. Similarly, the second similar case $C_2^d$ with better attributes (less credit amount, duration, and longer time living in the present residence, and more credits at the current bank) still defaults. Thus, to avoid the potential risk, the bank has a reason to reject similar cases. Meanwhile, two quality cases $C_3^n$ and $C_4^n$ can provide suggestions for the customer to improve his case and obtain a successful application. For the same duration, the less credit amount is important, and even there is a need to lessen the credit installments as a percentage of disposable income to a low level. Besides, considering the age, an important feature aforementioned, longer employment duration, and proof of property are important to decrease the expectation of the credit risk. 


\begin{table*}[ht!]
\aboverulesep=0.3ex 
\belowrulesep=0.3ex 
\caption{Features comparison for the south German credit risk.}\label{t:feature_example}
 \begin{threeparttable}
 \begin{tabular*}{\textwidth}{p{4cm} >{\centering\arraybackslash}p{2cm}: >{\centering\arraybackslash}p{2cm} >{\centering\arraybackslash}p{2cm} >{\centering\arraybackslash}p{2cm} :>{\centering\arraybackslash}p{2cm}} 
\toprule

Feature&$C_0$&$C_1^d$&$C_2^d$&$C_3^n$&$C_4^n$\\
\midrule
Status        &1    &1    &1    &1    &1\\
Duration      &11   &9    &5    &11   &11\\
Credit history&4    &4    &4    &4    &4\\
Purpose       &0    &0    &0    &0    &0\\
Amount        &3,905&2,799&3,676&3,499&691\\

Savings            &1    &1    &1    &1    &1\\
Employment duration&3    &3    &3    &3    &5\\
Installment rate   &2    &2    &1    &3    &4\\
Personal status    &3    &3    &3    &2    &3\\
Other debtors      &1    &1    &1    &2    &1\\

Present residence      &2     &2&3          &2&3 \\
Property               &1     &1&1          &1&2\\
Age                    &36    &36&37           &28&35\\
Other installment plans&3    &3&3        &3&3\\
Housing                &1    &1&1        &2&2\\

Number credits  &2   &2&3      &2&2\\
Job             &3   &3&3      &3&3\\
People liable   &1   &1&1      &2&2\\
Telephone       &1   &1&1      &1&1\\
Foreign worker  &2   &2&2      &2&2\\
\bottomrule
\end{tabular*}
 \begin{tablenotes}
 \scriptsize
      \item Notes: What the value of the features stands for can be found in Appendix \ref{german_data}. $C_0$ is the application case which has been predicted as that there is a possibility of a loss resulting from the applier's failure to meet contractual obligations. $C_1^d$ (1st), $C_2^d$ (2nd) and $C_3^n$ (3rd) are three most similar cases with $C_0$. $C_1^d$ and $C_2^d$ are two default cases. $C_3^n$ and $C_4^n$ are the most two similar quality cases. Superscript $d$ and $n$ stands for default and non-default cases, respectively.
    \end{tablenotes}
  \end{threeparttable}
\end{table*}
\subsubsection{Results explanation based on probability}
The application case $C_0$ is classified in terms of voting its three most similar cases. According to the log likelihood function introduced in section \ref{Probability}, we can obtain the probability weights of the most, second, third similar cases $C_1^d$, $C_2^d$ and $C_3^n$ are 0.4879, 0.3123 and 0.1998, respectively. Thus, there is 80.02\% probability that $C_0$ is a bad case. The bank has a high confidence to reject the application.  

\subsubsection{Results explanation based on feature relevance}
The global similarity is calculated in terms of the relevance scores of features. Table \ref{t:feature_imp} shows the different feature relevance when making a prediction for the risk of credit applications. From Table \ref{t:feature_imp}, we can observe that the financial status is the most important feature. The second is the duration that the customer wants to take the credit from a bank. The third and fourth ones are credit amount and age, respectively. In contrast, whether the applier is a foreign worker are not relatively important. Understanding the relevance of the features is significant for a bank to filter applications and make approval decisions.

\begin{centering}
\begin{figure}[ht!]
\centering
\includegraphics[width=\textwidth]{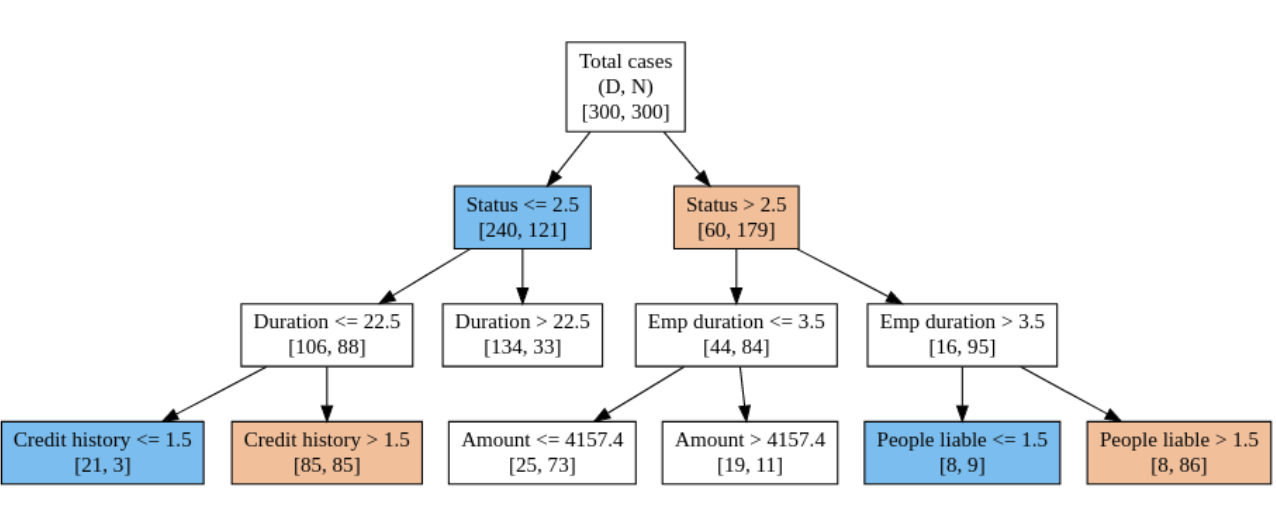}
\caption{ The most likely causes for a particular default based on CID tree scoring. The decision tree is built using the C4.5 algorithm, and the prominent features have been highlighted (node P (blue) and node S (yellow)).}
\label{fig:CID}
\end{figure} 
\end{centering}

\begin{table*}[ht!]
\caption{Feature relevance for the south German credit risk.}\label{t:feature_imp}
 \begin{threeparttable}
 \begin{tabular*}{\textwidth}{lP{3.8cm} l P{3.8cm}  } 
\toprule
Feature&Feature relevance (\%) & Feature&Feature relevance (\%)\\
\midrule
Status&13.84 & Present residence&4.91 \\
Duration& 7.01&Property&4.72\\
Credit history&5.65 &Age&6.07\\
Purpose&5.57 &Other installment plans&3.61 \\
Amount& 6.60 &Housing&3.88 \\

Savings&4.82&Number credits&3.57 \\
Employment duration&5.48&Job&4.34 \\
Installment rate& 4.93&People liable&2.82 \\
Personal status&  4.21&Telephone&3.33\\
Other debtors&3.25 &Foreign worker&1.38 \\
\bottomrule
\end{tabular*}
 \begin{tablenotes}
 \scriptsize
      \item Notes: The tables presents the relevance of each feature. The sum of all the value is equal to 100.
    \end{tablenotes}
  \end{threeparttable}
  
\end{table*}
As aforementioned, we use a CID tree to detect the most likely causes for a default case. The decision tree is built using the C4.5 algorithm, and the prominent features have been highlighted (node P (blue) and node S (yellow)), as shown in Figure \ref{fig:CID}, and the score for each node is calculated based on the equation \ref{cid_eq}, as shown in Table \ref{t:score_cid}. From Figure \ref{fig:CID} and Table \ref{t:score_cid}, we can observe that the inferior status of checking account, bad credit history, and too many people liable of credit applicants are the three most likely causes for their cases default. Compared with numeric features, such as the duration and amount, the three categorical and ordinal features perform better for discriminating default and non-default cases. Thus, a creditor is able to determine if an applicant is a good credit risk based on the following criterion:
\begin{itemize}
\item The status of the credit applicants' checking account with the bank is essential to be active with a positive balance.
\item Credit applicants with defective credit history have a high possibility of default again. 
\item The fewer the number of persons who financially depend on a credit applicant, the better. 
\end{itemize}
The CID tree provides complementary information for CBR system to detect the causes of default cases.

\begin{table*}[ht!]
\caption{The scores obtained from CID.}\label{t:score_cid}
 \begin{threeparttable}
 \begin{tabular*}{\textwidth}{lP{2.5cm} P{2.5cm}  P{2.5cm}  P{2.2cm} } 
\toprule
Feature&D &N& Score&1/Score\\
\midrule
Total cases&300 & 300&0 &0\\
\rowcolor{lightgray} Status $\leq$ 2.5&240 & 121&35.0 &0.0\\
\rowcolor{lightgray} Status > 2.5  &69& 179&0.0 &35.0\\
Duration $\leq$ 2.5&106 & 88&0.1 &11\\
Duration >22.5&134 & 33&11 &0.1\\
Employment duration $\leq$ 3.5&44 & 84&9.7 &0.1\\
Employment duration >3.5&16 & 95&0.1 &9.7\\
\rowcolor{lightgray} Credit history $\leq$ 1.5&21 & 3&81 &0.0\\
\rowcolor{lightgray} Credit history >1.5&85 & 85&0.0 &81\\
Amount $\leq$ 4157.4&25 & 73&0.0 &25.4\\
Amount >4157.4&19 & 11&25.4 &0.0\\
\rowcolor{lightgray} People liable $\leq$ 1.5&86 & 8&95.6 &0.0\\
\rowcolor{lightgray} People liable >1.5&8 & 86&0.0 &95.6\\
\bottomrule
\end{tabular*}
\end{threeparttable}
\end{table*}
\subsubsection{Further information detection}
According to the feature relevance analysis above, we know features have different contributions when determining if a credit applicant is qualified. Typically, status, credit history, and people liable are the three most likely decisive reasons, for all of which a higher value is better. In this section, we implement the clustering technique to extract more criteria for credit application decision-making by using group information from the three salient features. 

\begin{centering}
\begin{figure}[!ht]
\centering
\includegraphics[width=0.8\textwidth]{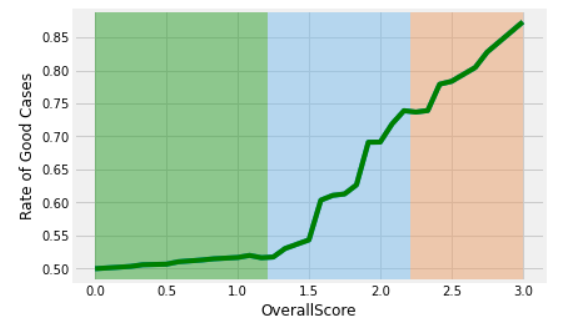}
\caption{\centering The rate of good cases increases with the overall score of cases increases. Green, blue, and red areas stand for the low, moderate, and high value groups, respectively.}
\label{fig:clustering}
\end{figure} 
\end{centering}

\begin{centering}
\begin{figure}[ht!]
\centering
\includegraphics[width=0.85\textwidth]{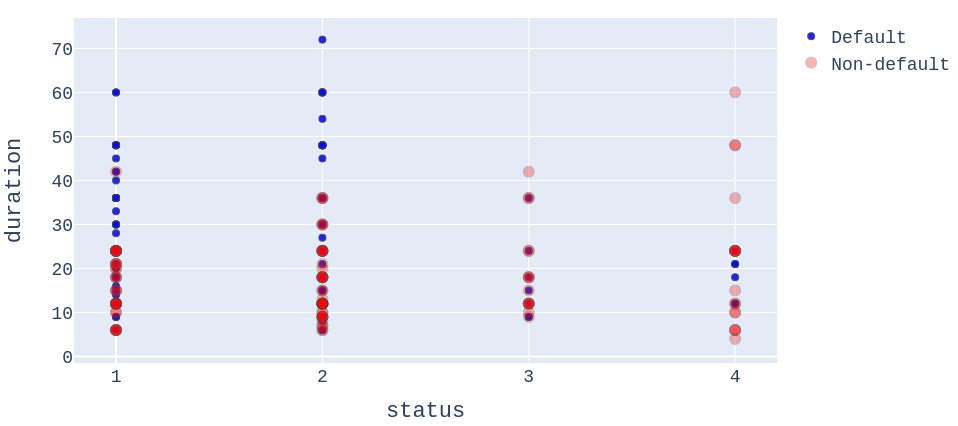}
\caption{\centering Scatterplot of duration and status of the cases data from the moderate group, with blue and red color denoting default and non-default, respectively. }
\label{fig:s_d}
\end{figure} 
\end{centering}

\begin{centering}
\begin{figure}[ht!]
\centering
\includegraphics[width=0.85\textwidth]{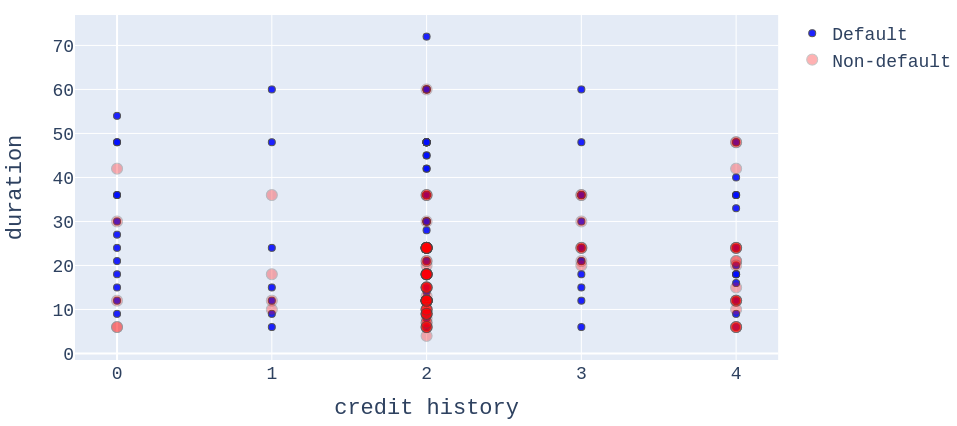}
\caption{\centering Scatterplot of duration and credit history of the cases data from the moderate group, with blue and red color denoting default and non-default, respectively.}
\label{fig:ch_d}
\end{figure} 
\end{centering}

\begin{centering}
\begin{figure}[ht!]
\centering
\includegraphics[width=0.85\textwidth]{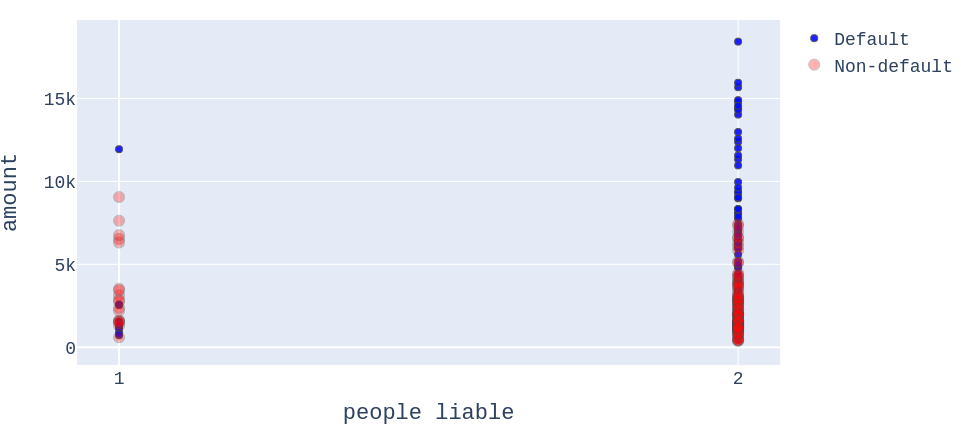}
\caption{\centering Scatterplot of amount and people liable of the cases data from the moderate group, with blue and red color denoting default and non-default, respectively.}
\label{fig:pl_a}
\end{figure} 
\end{centering}

Due to the features with varying degrees of magnitude and range, we normalized their value on a scale of 0 to 1. Consequently, the overall score of each case (sum value of the three features) ranges from 0 to 3. The relations between the overall score and the rate of good cases can be found in Figure \ref{fig:clustering}. In addition, the customer segmentation is useful in understanding demographic and psychographic profiles of the credit applicants in a bank \cite{Zakrzewska2005197}. We cluster the cases into three groups using k-means algorithm and highlight each group with different colors, shown in Figure \ref{fig:clustering}. From the figure, we can see that the non-default rate of the low value group is steadily less than 53\% while the moderate and high value group increases significantly. Compared to the low value group (high probability to default) and high value group (low probability to default), the moderate group has the high potential to increase the rate of good cases with cost-effective assists from the bank. The applicants in moderate group deserve the priority from bank to conduct group analysis and offer constructive advice to escalate the likelihood of their successful applications. To further detect information for credit application decision making, the scatterplots, duration and status, duration and credit history and amount and people liable of the cases data from the moderate group, are shown in Figure \ref{fig:s_d}, \ref{fig:ch_d} and \ref{fig:pl_a}, respectively. Blue color dots denote default cases, and red dots denote non-default. 

From Figure \ref{fig:s_d}, we can see that if the value of duration excesses some value, the majority of application cases are default when the status of checking accounts of customers are not active with a positive balance (status = 1 or 2). This can direct the bank by setting thresholds to effectively filter and review cases in a preliminary stage. In addition, reduction of credit application duration would be a constructive suggestion in a quantitative way for a particular customer to improve his application when his financial status is not competent.

From Figure \ref{fig:ch_d}, we can observe that it is meaningful to set a decisive threshold to reject credit application with a long duration requirement when the debtor has no history of credits taken or all credits paid back duly (credit history = 2). Meanwhile, if the debtor has a delayed history of paying off or a critical account elsewhere, he has a high probability of defaulting with any duration magnitude. There are no obvious relations between the duration and creditability of a debtor when he has all credits at this bank paid back duly. It is interesting to note that if the credit applicant has existing credits paid back duly till now, the short duration is not a good signal. The present applied credit would be abused for repayment.  

In Figure \ref{fig:pl_a}, we can see that the majority of the credit applicants in the moderate group have less than 2 people who financially depend on them. And, it is obvious that there is a decision threshold for approving credit, considered in terms of amount when 0 to 2 people financially supported by the applicant. 


\section{Conclusion}
Financial risks are uncertainties associated with financial decisions, such as credit application approval and bank customers' churn reduction. In recent years, some complex black-box AI methods have achieved unprecedented levels of performance when learning to solve increasing complex computational tasks, including financial risk detection. However, the GDPR rule in Europe contests any automated decision-making that was made on a solely algorithmic basis. Additionally, decision-making processes are required to be accompanied by a meaningful explanation. Consequently, financial industries guided by the regulation urge the need for innovative research on XAI methods. CBR system is an XAI method, which has been identified as a useful method in real-life applications. 

In this study, a data-driven explainable CBR system is proposed for solving financial risk prediction problems. In particular, feature relevance scoring methods are applied for assigning global similarity weights for attributes, and the PSO algorithm is employed for optimizing the parameters of local similarity functions. The proposed data-driven approach provides a way to overcome the drawback of the standard CBR system, which highly depends on domain knowledge and prior experience when building a successful model. The experimental results show that the proposed CBR method has a relatively superior prediction performance compared to the widely used classification machine learning methods. In addition, compared to other black-box machine learning methods, the CBR system is capable of interpreting the result of the financial risk prediction, and further detects the schemes to decrease or avoid financial risk. This characteristic is significantly helpful for both financial institutions and their customers. In particular, we introduce four major explanation goals and conduct an experiment to outline a unified view on explanation in the CBR system, using a German credit risk dataset. The results show the scheme to explain the decision making process based on the CBR system and explore more decisive information for the bank.

For future studies, several extensions of the current study can be developed. In the proposed CBR system, the design of global similarity depends on the existing feature scoring methods. In further research, we will explore a general way to detect the optimal feature weights. Furthermore, the main limitation of the proposed CBR system is that its training is extremely time-consuming. It is because the PSO algorithm is a computationally-intensive optimization method to search for the parameters of local similarity functions. More efficient optimization algorithms for creating the CBR system are needed. Finally, the study was carried out using the five financial risk datasets, but the generality of the proposed CBR system ensures a possible application to other decision-support systems.

\section*{Acknowledgment}
This work acknowledges research support by COST Action “Fintech and Artificial Intelligence in Finance - Towards a transparent financial industry” (FinAI) CA19130. Critical comments and advice from Denis M. Becker, Christian Ewald, Wolfgang Karl Härdle, Steven Ongena and Endre Jo Reite are gratefully acknowledged. The computations were performed on resources provided by UNINETT Sigma2 - the National Infrastructure for High Performance Computing and Data Storage in Norway.

\newpage 


\bibliographystyle{abbrvnat} 
\bibliography{literature.bib} 

\begin{thebibliography}{74}
\providecommand{\natexlab}[1]{#1}
\providecommand{\url}[1]{\texttt{#1}}
\expandafter\ifx\csname urlstyle\endcsname\relax
  \providecommand{\doi}[1]{doi: #1}\else
  \providecommand{\doi}{doi: \begingroup \urlstyle{rm}\Url}\fi

\bibitem[Aamodt and Plaza(1994)]{Aamodt1994}
A.~Aamodt and E.~Plaza.
\newblock Case-based reasoning: Foundational issues, methodological variations,
  and system approaches.
\newblock \emph{AI Communications}, 7:\penalty0 39--59, 1994.
\newblock \doi{10.3233/AIC-1994-7104}.
\newblock 1.

\bibitem[Aamodt et~al.(1998)Aamodt, Sandtorv, and
  Winnem]{Aamodt98combiningcase}
A.~Aamodt, H.~A. Sandtorv, and O.~M. Winnem.
\newblock Combining case based reasoning and data mining - a way of revealing
  and reusing rams experience.
\newblock In \emph{In Lydersen, Hansen, Sandtorv (eds.), Safety and
  Reliability; Proceedings of ESREL ’98}, pages 16--19, 1998.

\bibitem[Abdou et~al.(2008)Abdou, Pointon, and El-Masry]{ABDOU20081275}
H.~Abdou, J.~Pointon, and A.~El-Masry.
\newblock Neural nets versus conventional techniques in credit scoring in
  egyptian banking.
\newblock \emph{Expert Systems with Applications}, 35\penalty0 (3):\penalty0
  1275--1292, 2008.
\newblock ISSN 0957-4174.
\newblock \doi{https://doi.org/10.1016/j.eswa.2007.08.030}.

\bibitem[Ahn and jae Kim(2009)]{AHN2009599}
H.~Ahn and K.~jae Kim.
\newblock Bankruptcy prediction modeling with hybrid case-based reasoning and
  genetic algorithms approach.
\newblock \emph{Applied Soft Computing}, 9\penalty0 (2):\penalty0 599--607,
  2009.
\newblock ISSN 1568-4946.
\newblock \doi{https://doi.org/10.1016/j.asoc.2008.08.002}.

\bibitem[Alam et~al.(2020)Alam, Shaukat, Hameed, Luo, Sarwar, Shabbir, Li, and
  Khushi]{Alam2020201173}
T.~M. Alam, K.~Shaukat, I.~A. Hameed, S.~Luo, M.~U. Sarwar, S.~Shabbir, J.~Li,
  and M.~Khushi.
\newblock An investigation of credit card default prediction in the imbalanced
  datasets.
\newblock \emph{IEEE Access}, 8:\penalty0 201173--201198, 2020.
\newblock \doi{10.1109/ACCESS.2020.3033784}.

\bibitem[{Arshadi} and {Jurisica}(2005)]{Arshadi20051127}
N.~{Arshadi} and I.~{Jurisica}.
\newblock Data mining for case-based reasoning in high-dimensional biological
  domains.
\newblock \emph{IEEE Transactions on Knowledge and Data Engineering},
  17\penalty0 (8):\penalty0 1127--1137, 2005.
\newblock \doi{10.1109/TKDE.2005.124}.

\bibitem[Atiya(2005)]{Atiya2005731}
A.~F. Atiya.
\newblock {Estimating the Posterior Probabilities Using the K-Nearest Neighbor
  Rule}.
\newblock \emph{Neural Computation}, 17\penalty0 (3):\penalty0 731--740, 03
  2005.
\newblock ISSN 0899-7667.
\newblock \doi{10.1162/0899766053019971}.

\bibitem[Bach and Althoff(2012)]{Bach201217}
K.~Bach and K.-D. Althoff.
\newblock Developing case-based reasoning applications using mycbr 3.
\newblock In B.~D. Agudo and I.~Watson, editors, \emph{Case-Based Reasoning
  Research and Development}, pages 17--31, Berlin, Heidelberg, 2012. Springer
  Berlin Heidelberg.
\newblock ISBN 978-3-642-32986-9.

\bibitem[Bensic et~al.(2005)Bensic, Sarlija, and Zekic-Susac]{Bensic2005133}
M.~Bensic, N.~Sarlija, and M.~Zekic-Susac.
\newblock Modelling small-business credit scoring by using logistic regression,
  neural networks and decision trees.
\newblock \emph{Intelligent Systems in Accounting, Finance and Management},
  13\penalty0 (3):\penalty0 133--150, 2005.
\newblock \doi{https://doi.org/10.1002/isaf.261}.

\bibitem[Brown and Gupta(1994)]{Brown1994}
C.~E. Brown and U.~G. Gupta.
\newblock Applying case-based reasoning to the accounting domain.
\newblock \emph{Intelligent Systems in Accounting, Finance and Management},
  3\penalty0 (3):\penalty0 205--221, 1994.
\newblock \doi{10.1002/j.1099-1174.1994.tb00066.x}.

\bibitem[{Brunette} et~al.(2009){Brunette}, {Flemmer}, and
  {Flemmer}]{Brunette2009}
E.~S. {Brunette}, R.~C. {Flemmer}, and C.~L. {Flemmer}.
\newblock A review of artificial intelligence.
\newblock In \emph{2009 4th International Conference on Autonomous Robots and
  Agents}, pages 385--392, 2009.
\newblock \doi{10.1109/ICARA.2000.4804025}.

\bibitem[Bryant(1997)]{Bryant1997}
S.~M. Bryant.
\newblock A case-based reasoning approach to bankruptcy prediction modeling.
\newblock \emph{Intelligent Systems in Accounting, Finance and Management},
  6\penalty0 (3):\penalty0 195--214, 1997.

\bibitem[Ceriani and Verme(2012)]{Ceriani2012}
L.~Ceriani and P.~Verme.
\newblock The origins of the gini index: extracts from variabilit{\`a} e
  mutabilit{\`a} (1912) by corrado gini.
\newblock \emph{The Journal of Economic Inequality}, 10\penalty0 (3):\penalty0
  421--443, Sep 2012.
\newblock ISSN 1573-8701.
\newblock \doi{10.1007/s10888-011-9188-x}.

\bibitem[Chen et~al.(2011)Chen, Härdle, and Moro]{Chen2011135}
S.~Chen, W.~K. Härdle, and R.~A. Moro.
\newblock Modeling default risk with support vector machines.
\newblock \emph{Quantitative. Finance}, 11\penalty0 (1):\penalty0 135--154,
  2011.
\newblock \doi{10.1080/14697680903410015}.

\bibitem[Chi et~al.(1993)Chi, Chen, and Kiang]{CHI199367}
R.~T. Chi, M.~Chen, and M.~Y. Kiang.
\newblock Generalized case-based reasoning system for portfolio management.
\newblock \emph{Expert Systems with Applications}, 6\penalty0 (1):\penalty0 67
  -- 76, 1993.
\newblock \doi{https://doi.org/10.1016/0957-4174(93)90019-3}.
\newblock Special Issue: Case-Based Reasoning and its Applications.

\bibitem[Cost and Salzberg(1993)]{Cost1993}
S.~Cost and S.~Salzberg.
\newblock A weighted nearest neighbor algorithm for learning with symbolic
  features.
\newblock \emph{Machine Learning}, 10\penalty0 (1):\penalty0 57--78, Jan 1993.
\newblock ISSN 1573-0565.
\newblock \doi{10.1023/A:1022664626993}.

\bibitem[Cunningham et~al.(2003)Cunningham, Doyle, and
  Loughrey]{Cunningham2003122}
P.~Cunningham, D.~Doyle, and J.~Loughrey.
\newblock An evaluation of the usefulness of case-based explanation.
\newblock In K.~D. Ashley and D.~G. Bridge, editors, \emph{Case-Based Reasoning
  Research and Development}, pages 122--130, Berlin, Heidelberg, 2003. Springer
  Berlin Heidelberg.
\newblock ISBN 978-3-540-45006-1.

\bibitem[{Dal Pozzolo} et~al.(2014){Dal Pozzolo}, Caelen, {Le Borgne},
  Waterschoot, and Bontempi]{DALPOZZOLO20144915}
A.~{Dal Pozzolo}, O.~Caelen, Y.-A. {Le Borgne}, S.~Waterschoot, and
  G.~Bontempi.
\newblock Learned lessons in credit card fraud detection from a practitioner
  perspective.
\newblock \emph{Expert Systems with Applications}, 41\penalty0 (10):\penalty0
  4915--4928, 2014.
\newblock ISSN 0957-4174.
\newblock \doi{https://doi.org/10.1016/j.eswa.2014.02.026}.
\newblock URL
  \url{https://www.sciencedirect.com/science/article/pii/S095741741400089X}.

\bibitem[Ebrahimi(2017)]{Ebrahimi2017}
Ebrahimi.
\newblock {Kaggle Financial Distress Prediction}.
\newblock \url{https://www.kaggle.com/shebrahimi/financial-distress}, December
  2017.

\bibitem[Fayyad et~al.(1996)Fayyad, Piatetsky-Shapiro, and Smyth]{Fayyad199637}
U.~Fayyad, G.~Piatetsky-Shapiro, and P.~Smyth.
\newblock From data mining to knowledge discovery in databases.
\newblock \emph{AI Magazine}, 17\penalty0 (3):\penalty0 37, Mar. 1996.
\newblock \doi{10.1609/aimag.v17i3.1230}.

\bibitem[Gardner and Dorling(1998)]{GARDNER19982627}
M.~Gardner and S.~Dorling.
\newblock Artificial neural networks (the multilayer perceptron)—a review of
  applications in the atmospheric sciences.
\newblock \emph{Atmospheric Environment}, 32\penalty0 (14):\penalty0
  2627--2636, 1998.
\newblock ISSN 1352-2310.
\newblock \doi{https://doi.org/10.1016/S1352-2310(97)00447-0}.
\newblock URL
  \url{https://www.sciencedirect.com/science/article/pii/S1352231097004470}.

\bibitem[{Gouttaya} and {Begdouri}(2012)]{Gouttaya2012136}
N.~{Gouttaya} and A.~{Begdouri}.
\newblock Integrating data mining with case based reasoning (cbr) to improve
  the proactivity of pervasive applications.
\newblock In \emph{2012 Colloquium in Information Science and Technology},
  pages 136--141, 2012.
\newblock \doi{10.1109/CIST.2012.6388077}.

\bibitem[Grömping(2019)]{Ulrike2019}
U.~Grömping.
\newblock \emph{South German Credit Data: Correcting a Widely Used Data Set}.
\newblock Beuth University of Applied Sciences Berlin, 2019.

\bibitem[Guessoum et~al.(2014)Guessoum, Laskri, and Lieber]{GUESSOUM2014267}
S.~Guessoum, M.~T. Laskri, and J.~Lieber.
\newblock Respidiag: A case-based reasoning system for the diagnosis of chronic
  obstructive pulmonary disease.
\newblock \emph{Expert Systems with Applications}, 41\penalty0 (2):\penalty0
  267--273, 2014.
\newblock ISSN 0957-4174.
\newblock \doi{https://doi.org/10.1016/j.eswa.2013.05.065}.

\bibitem[Ha et~al.(2019)Ha, Lu, Choi, Nguyen, and Yoon]{Ha2019511}
V.-S. Ha, D.-N. Lu, G.~S. Choi, H.-N. Nguyen, and B.~Yoon.
\newblock Improving credit risk prediction in online peer-to-peer (p2p) lending
  using feature selection with deep learning.
\newblock In \emph{2019 21st International Conference on Advanced Communication
  Technology (ICACT)}, pages 511--515, 2019.
\newblock \doi{10.23919/ICACT.2019.8701943}.

\bibitem[Henley and Hand(1996)]{Henley199677}
W.~E. Henley and D.~J. Hand.
\newblock A k-nearest-neighbour classifier for assessing consumer credit risk.
\newblock \emph{Journal of the Royal Statistical Society: Series D (The
  Statistician)}, 45\penalty0 (1):\penalty0 77--95, 1996.
\newblock \doi{https://doi.org/10.2307/2348414}.

\bibitem[Hu et~al.(2016)Hu, Xia, Skitmore, and Chen]{HU201665}
X.~Hu, B.~Xia, M.~Skitmore, and Q.~Chen.
\newblock The application of case-based reasoning in construction management
  research: An overview.
\newblock \emph{Automation in Construction}, 72:\penalty0 65--74, 2016.
\newblock ISSN 0926-5805.
\newblock \doi{https://doi.org/10.1016/j.autcon.2016.08.023}.

\bibitem[Hwang and Chu(2018)]{Hwang2018419}
R.-C. Hwang and C.-K. Chu.
\newblock A logistic regression point of view toward loss given default
  distribution estimation.
\newblock \emph{Quantitative Finance}, 18\penalty0 (3):\penalty0 419--435,
  2018.
\newblock \doi{10.1080/14697688.2017.1310393}.

\bibitem[Ince(2014)]{INCE2014205}
H.~Ince.
\newblock Short term stock selection with case-based reasoning technique.
\newblock \emph{Applied Soft Computing}, 22:\penalty0 205--212, 2014.
\newblock ISSN 1568-4946.
\newblock \doi{https://doi.org/10.1016/j.asoc.2014.05.017}.

\bibitem[Jaiswal and Bach(2019)]{Jaiswal2019}
A.~Jaiswal and K.~Bach.
\newblock A data-driven approach for determining weights in global similarity
  functions.
\newblock In K.~Bach and C.~Marling, editors, \emph{Case-Based Reasoning
  Research and Development}, pages 125--139, Cham, 2019. Springer International
  Publishing.

\bibitem[Kao et~al.(2012)Kao, Chiu, and Chiu]{KAO2012245}
L.-J. Kao, C.-C. Chiu, and F.-Y. Chiu.
\newblock A bayesian latent variable model with classification and regression
  tree approach for behavior and credit scoring.
\newblock \emph{Knowledge-Based Systems}, 36:\penalty0 245--252, 2012.
\newblock ISSN 0950-7051.
\newblock \doi{https://doi.org/10.1016/j.knosys.2012.07.004}.

\bibitem[Kleinbaum(1994)]{Kleinbaum1994}
D.~G. Kleinbaum.
\newblock \emph{Introduction to Logistic Regression}, pages 1--38.
\newblock Springer New York, New York, NY, 1994.
\newblock ISBN 978-1-4757-4108-7.
\newblock \doi{10.1007/978-1-4757-4108-7_1}.
\newblock URL \url{https://doi.org/10.1007/978-1-4757-4108-7_1}.

\bibitem[Kononenko et~al.(1997)Kononenko, {\v{S}}imec, and
  Robnik-{\v{S}}ikonja]{Kononenko1997}
I.~Kononenko, E.~{\v{S}}imec, and M.~Robnik-{\v{S}}ikonja.
\newblock Overcoming the myopia of inductive learning algorithms with relieff.
\newblock \emph{Applied Intelligence}, 7\penalty0 (1):\penalty0 39--55, Jan
  1997.
\newblock ISSN 1573-7497.
\newblock \doi{10.1023/A:1008280620621}.

\bibitem[Kraskov et~al.(2004)Kraskov, St\"ogbauer, and
  Grassberger]{Kraskov2004}
A.~Kraskov, H.~St\"ogbauer, and P.~Grassberger.
\newblock Estimating mutual information.
\newblock \emph{Phys. Rev. E}, 69:\penalty0 066138, Jun 2004.
\newblock \doi{10.1103/PhysRevE.69.066138}.

\bibitem[Kullback(1959)]{Kullback59}
S.~Kullback.
\newblock \emph{Information Theory and Statistics}.
\newblock Wiley, New York, 1959.

\bibitem[Lahmiri and Bekiros(2019)]{Salim20191569}
S.~Lahmiri and S.~Bekiros.
\newblock Can machine learning approaches predict corporate bankruptcy?
  evidence from a qualitative experimental design.
\newblock \emph{Quantitative Finance}, 19\penalty0 (9):\penalty0 1569--1577,
  2019.
\newblock \doi{10.1080/14697688.2019.1588468}.

\bibitem[Lamy et~al.(2019)Lamy, Sekar, Guezennec, Bouaud, and
  Séroussi]{LAMY201942}
J.-B. Lamy, B.~Sekar, G.~Guezennec, J.~Bouaud, and B.~Séroussi.
\newblock Explainable artificial intelligence for breast cancer: A visual
  case-based reasoning approach.
\newblock \emph{Artificial Intelligence in Medicine}, 94:\penalty0 42--53,
  2019.
\newblock ISSN 0933-3657.
\newblock \doi{https://doi.org/10.1016/j.artmed.2019.01.001}.

\bibitem[Li and Sun(2010)]{LI2010137}
H.~Li and J.~Sun.
\newblock Business failure prediction using hybrid2 case-based reasoning
  (h2cbr).
\newblock \emph{Computers \& Operations Research}, 37\penalty0 (1):\penalty0
  137--151, 2010.
\newblock ISSN 0305-0548.
\newblock \doi{https://doi.org/10.1016/j.cor.2009.04.003}.

\bibitem[Li and Hand(2002)]{Li2002647}
H.~G. Li and D.~J. Hand.
\newblock Direct versus indirect credit scoring classifications.
\newblock \emph{Journal of the Operational Research Society}, 53\penalty0
  (6):\penalty0 647--654, 2002.
\newblock \doi{10.1057/palgrave.jors.2601346}.

\bibitem[Li et~al.(2018)Li, Jamieson, DeSalvo, Rostamizadeh, and
  Talwalkar]{JMLR:v18:16-558}
L.~Li, K.~Jamieson, G.~DeSalvo, A.~Rostamizadeh, and A.~Talwalkar.
\newblock Hyperband: A novel bandit-based approach to hyperparameter
  optimization.
\newblock \emph{Journal of Machine Learning Research}, 18\penalty0
  (185):\penalty0 1--52, 2018.
\newblock URL \url{http://jmlr.org/papers/v18/16-558.html}.

\bibitem[Likas et~al.(2003)Likas, Vlassis, and {J. Verbeek}]{LIKAS2003451}
A.~Likas, N.~Vlassis, and J.~{J. Verbeek}.
\newblock The global k-means clustering algorithm.
\newblock \emph{Pattern Recognition}, 36\penalty0 (2):\penalty0 451--461, 2003.
\newblock ISSN 0031-3203.
\newblock \doi{https://doi.org/10.1016/S0031-3203(02)00060-2}.
\newblock Biometrics.

\bibitem[Lin and Ding(2011)]{LIN201164}
H.~Lin and H.~Ding.
\newblock Predicting ion channels and their types by the dipeptide mode of
  pseudo amino acid composition.
\newblock \emph{Journal of Theoretical Biology}, 269\penalty0 (1):\penalty0 64
  -- 69, 2011.
\newblock ISSN 0022-5193.
\newblock \doi{https://doi.org/10.1016/j.jtbi.2010.10.019}.

\bibitem[Mitchell(1997)]{Mitchell1997}
T.~M. Mitchell.
\newblock \emph{Machine Learning}.
\newblock McGraw-Hill, Inc., USA, 1 edition, 1997.
\newblock ISBN 0070428077.

\bibitem[Mohammed et~al.(2018)Mohammed, {Abd Ghani}, Arunkumar, Obaid, Mostafa,
  Jaber, Burhanuddin, Matar, khalid abdullatif, and Ibrahim]{MOHAMMED2018212}
M.~A. Mohammed, M.~K. {Abd Ghani}, N.~Arunkumar, O.~I. Obaid, S.~A. Mostafa,
  M.~M. Jaber, M.~Burhanuddin, B.~M. Matar, S.~khalid abdullatif, and D.~A.
  Ibrahim.
\newblock Genetic case-based reasoning for improved mobile phone faults
  diagnosis.
\newblock \emph{Computers \& Electrical Engineering}, 71:\penalty0 212--222,
  2018.
\newblock ISSN 0045-7906.
\newblock \doi{https://doi.org/10.1016/j.compeleceng.2018.07.053}.

\bibitem[Morris(1994)]{Morris1994}
B.~W. Morris.
\newblock Scan: A case-based reasoning model for generating information system
  control recommendations.
\newblock \emph{Intelligent Systems in Accounting, Finance and Management},
  3\penalty0 (1):\penalty0 47--63, 1994.
\newblock \doi{10.1002/j.1099-1174.1994.tb00054.x}.

\bibitem[Moxey et~al.(2010)Moxey, Robertson, Newby, Hains, Williamson, and
  Pearson]{Moxey201025}
A.~Moxey, J.~Robertson, D.~Newby, I.~Hains, M.~Williamson, and S.-A. Pearson.
\newblock {Computerized clinical decision support for prescribing: provision
  does not guarantee uptake}.
\newblock \emph{Journal of the American Medical Informatics Association},
  17\penalty0 (1):\penalty0 25--33, 01 2010.
\newblock ISSN 1067-5027.
\newblock \doi{10.1197/jamia.M3170}.

\bibitem[Novaković(2011)]{Novakovic2011}
J.~Novaković.
\newblock Toward optimal feature selection using ranking methods and
  classification algorithms.
\newblock \emph{Yugoslav Journal of Operations Research}, 21\penalty0 (1),
  2011.
\newblock ISSN 2334-6043.
\newblock \doi{https://doi.org/0.2298/YJOR1101119N}.

\bibitem[O'Roarty et~al.(1997)O'Roarty, Patterson, McGreal, and
  Adair]{OROARTY1997417}
B.~O'Roarty, D.~Patterson, S.~McGreal, and A.~Adair.
\newblock A case-based reasoning approach to the selection of comparable
  evidence for retail rent determination.
\newblock \emph{Expert Systems with Applications}, 12\penalty0 (4):\penalty0
  417 -- 428, 1997.
\newblock \doi{https://doi.org/10.1016/S0957-4174(97)83769-4}.

\bibitem[Pavlidis et~al.(2012)Pavlidis, Tasoulis, Adams, and
  Hand]{Pavlidis20121645}
N.~G. Pavlidis, D.~K. Tasoulis, N.~M. Adams, and D.~J. Hand.
\newblock Adaptive consumer credit classification.
\newblock \emph{Journal of the Operational Research Society}, 63\penalty0
  (12):\penalty0 1645--1654, 2012.
\newblock \doi{10.1057/jors.2012.15}.

\bibitem[Peng et~al.(2011)Peng, Wang, Kou, and Shi]{PENG20112906}
Y.~Peng, G.~Wang, G.~Kou, and Y.~Shi.
\newblock An empirical study of classification algorithm evaluation for
  financial risk prediction.
\newblock \emph{Applied Soft Computing}, 11\penalty0 (2):\penalty0 2906 --
  2915, 2011.
\newblock ISSN 1568-4946.
\newblock \doi{https://doi.org/10.1016/j.asoc.2010.11.028}.

\bibitem[{Prati}(2012)]{Prati2012}
R.~C. {Prati}.
\newblock Combining feature ranking algorithms through rank aggregation.
\newblock In \emph{The 2012 International Joint Conference on Neural Networks
  (IJCNN)}, pages 1--8, 2012.
\newblock \doi{10.1109/IJCNN.2012.6252467}.

\bibitem[Quinlan(1993)]{Quinlan1993}
J.~R. Quinlan.
\newblock \emph{C4.5: Programs for Machine Learning}.
\newblock Morgan Kaufmann Publishers Inc., San Francisco, CA, USA, 1993.
\newblock ISBN 1558602380.

\bibitem[Rahman and Kumar(2020)]{Rahman2020}
M.~Rahman and V.~Kumar.
\newblock Machine learning based customer churn prediction in banking.
\newblock In \emph{2020 4th International Conference on Electronics,
  Communication and Aerospace Technology (ICECA)}, pages 1196--1201, 2020.
\newblock \doi{10.1109/ICECA49313.2020.9297529}.

\bibitem[Richter and Weber(2013)]{Michael2013}
M.~M. Richter and R.~O. Weber.
\newblock \emph{Case-Based Reasoning: A Textbook}.
\newblock Springer Publishing Company, Incorporated, 2013.
\newblock ISBN 364240166X.

\bibitem[Sariev and Germano(2020)]{Eduard2020311}
E.~Sariev and G.~Germano.
\newblock Bayesian regularized artificial neural networks for the estimation of
  the probability of default.
\newblock \emph{Quantitative Finance}, 20\penalty0 (2):\penalty0 311--328,
  2020.
\newblock \doi{10.1080/14697688.2019.1633014}.

\bibitem[Selvamani and Khemani(2005)]{Radhika2005786}
B.~R. Selvamani and D.~Khemani.
\newblock Decision tree induction with cbr.
\newblock In S.~K. Pal, S.~Bandyopadhyay, and S.~Biswas, editors, \emph{Pattern
  Recognition and Machine Intelligence}, pages 786--791, Berlin, Heidelberg,
  2005. Springer Berlin Heidelberg.
\newblock ISBN 978-3-540-32420-1.

\bibitem[Sermpinis et~al.(2018)Sermpinis, Tsoukas, and Zhang]{SERMPINIS201819}
G.~Sermpinis, S.~Tsoukas, and P.~Zhang.
\newblock Modelling market implied ratings using lasso variable selection
  techniques.
\newblock \emph{Journal of Empirical Finance}, 48:\penalty0 19--35, 2018.
\newblock ISSN 0927-5398.
\newblock \doi{https://doi.org/10.1016/j.jempfin.2018.05.001}.
\newblock URL
  \url{https://www.sciencedirect.com/science/article/pii/S0927539818300318}.

\bibitem[Shin et~al.(1997)Shin, Shin, and Han]{Shin1997}
K.~Shin, T.~Shin, and I.~Han.
\newblock Using induction techniques to support case-based reasoning: a case of
  corporate bond rating.
\newblock \emph{Proceedings of the MS/OR society conference}, pages 199 -- 202,
  1997.

\bibitem[Song and Peng(2019)]{Song201984897}
Y.~Song and Y.~Peng.
\newblock A mcdm-based evaluation approach for imbalanced classification
  methods in financial risk prediction.
\newblock \emph{IEEE Access}, 7:\penalty0 84897--84906, 2019.
\newblock ISSN 2169-3536.
\newblock \doi{10.1109/ACCESS.2019.2924923}.

\bibitem[Song and Lu(2015)]{Song2015}
Y.-Y. Song and Y.~Lu.
\newblock Decision tree methods: applications for classification and
  prediction.
\newblock \emph{Shanghai archives of psychiatry}, 27\penalty0 (2):\penalty0
  130--135, Apr 2015.
\newblock ISSN 1002-0829.
\newblock URL \url{https://pubmed.ncbi.nlm.nih.gov/26120265}.

\bibitem[S{\o}rmo and Cassens(2004)]{Sormo2004165}
F.~S{\o}rmo and J.~Cassens.
\newblock Explanation goals in case-based reasoning.
\newblock \emph{In: Funk, P., González Calero, P.A. (eds.) ECCBR 2004. LNCS
  (LNAI)}, page 165–174, 2004.

\bibitem[S{\o}rmo et~al.(2005)S{\o}rmo, Cassens, and Aamodt]{Sormo2005}
F.~S{\o}rmo, J.~Cassens, and A.~Aamodt.
\newblock Explanation in case-based reasoning--perspectives and goals.
\newblock \emph{Artificial Intelligence Review}, 24\penalty0 (2):\penalty0
  109--143, Oct 2005.
\newblock ISSN 1573-7462.
\newblock \doi{10.1007/s10462-005-4607-7}.
\newblock URL \url{https://doi.org/10.1007/s10462-005-4607-7}.

\bibitem[Stevenson et~al.(2021)Stevenson, Mues, and Bravo]{STEVENSON2021758}
M.~Stevenson, C.~Mues, and C.~Bravo.
\newblock The value of text for small business default prediction: A deep
  learning approach.
\newblock \emph{European Journal of Operational Research}, 295\penalty0
  (2):\penalty0 758--771, 2021.
\newblock ISSN 0377-2217.
\newblock \doi{https://doi.org/10.1016/j.ejor.2021.03.008}.

\bibitem[Stoltzfus(2011)]{Stoltzfus20111099}
J.~C. Stoltzfus.
\newblock Logistic regression: A brief primer.
\newblock \emph{Academic Emergency Medicine}, 18\penalty0 (10):\penalty0
  1099--1104, 2011.
\newblock \doi{https://doi.org/10.1111/j.1553-2712.2011.01185.x}.
\newblock URL
  \url{https://onlinelibrary.wiley.com/doi/abs/10.1111/j.1553-2712.2011.01185.x}.

\bibitem[Tibshirani(1996)]{Robert1996267}
R.~Tibshirani.
\newblock Regression shrinkage and selection via the lasso.
\newblock \emph{Journal of the Royal Statistical Society: Series B
  (Methodological)}, 58\penalty0 (1):\penalty0 267--288, 1996.
\newblock \doi{https://doi.org/10.1111/j.2517-6161.1996.tb02080.x}.
\newblock URL
  \url{https://rss.onlinelibrary.wiley.com/doi/abs/10.1111/j.2517-6161.1996.tb02080.x}.

\bibitem[Trivedi(2020)]{TRIVEDI2020101413}
S.~K. Trivedi.
\newblock A study on credit scoring modeling with different feature selection
  and machine learning approaches.
\newblock \emph{Technology in Society}, 63:\penalty0 101413, 2020.
\newblock ISSN 0160-791X.
\newblock \doi{https://doi.org/10.1016/j.techsoc.2020.101413}.

\bibitem[Tsai and Wu(2008)]{TSAI20082639}
C.-F. Tsai and J.-W. Wu.
\newblock Using neural network ensembles for bankruptcy prediction and credit
  scoring.
\newblock \emph{Expert Systems with Applications}, 34\penalty0 (4):\penalty0
  2639--2649, 2008.
\newblock ISSN 0957-4174.
\newblock \doi{https://doi.org/10.1016/j.eswa.2007.05.019}.

\bibitem[Voigt and Bussche(2017)]{Voigt2017}
P.~Voigt and A.~v.~d. Bussche.
\newblock \emph{The EU General Data Protection Regulation (GDPR): A Practical
  Guide}.
\newblock Springer Publishing Company, Incorporated, 1st edition, 2017.
\newblock ISBN 3319579584.

\bibitem[Vukovic et~al.(2012)Vukovic, Delibasic, Uzelac, and
  Suknovic]{VUKOVIC20128389}
S.~Vukovic, B.~Delibasic, A.~Uzelac, and M.~Suknovic.
\newblock A case-based reasoning model that uses preference theory functions
  for credit scoring.
\newblock \emph{Expert Systems with Applications}, 39\penalty0 (9):\penalty0
  8389--8395, 2012.
\newblock ISSN 0957-4174.
\newblock \doi{https://doi.org/10.1016/j.eswa.2012.01.181}.

\bibitem[West(2000)]{WEST20001131}
D.~West.
\newblock Neural network credit scoring models.
\newblock \emph{Computers \& Operations Research}, 27\penalty0 (11):\penalty0
  1131--1152, 2000.
\newblock ISSN 0305-0548.
\newblock \doi{https://doi.org/10.1016/S0305-0548(99)00149-5}.

\bibitem[Wu and Lin(2005)]{10.1145/1089551.1089610}
J.~Wu and Z.~Lin.
\newblock Research on customer segmentation model by clustering.
\newblock In \emph{Proceedings of the 7th International Conference on
  Electronic Commerce}, ICEC '05, page 316–318, New York, NY, USA, 2005.
  Association for Computing Machinery.
\newblock ISBN 1595931120.
\newblock \doi{10.1145/1089551.1089610}.
\newblock URL \url{https://doi.org/10.1145/1089551.1089610}.

\bibitem[Yeh and hui Lien(2009)]{YEH20092473}
I.-C. Yeh and C.~hui Lien.
\newblock The comparisons of data mining techniques for the predictive accuracy
  of probability of default of credit card clients.
\newblock \emph{Expert Systems with Applications}, 36\penalty0 (2, Part
  1):\penalty0 2473--2480, 2009.
\newblock ISSN 0957-4174.
\newblock \doi{https://doi.org/10.1016/j.eswa.2007.12.020}.
\newblock URL
  \url{https://www.sciencedirect.com/science/article/pii/S0957417407006719}.

\bibitem[Zakrzewska and Murlewski(2005)]{Zakrzewska2005197}
D.~Zakrzewska and J.~Murlewski.
\newblock Clustering algorithms for bank customer segmentation.
\newblock In \emph{5th International Conference on Intelligent Systems Design
  and Applications (ISDA'05)}, pages 197--202, 2005.
\newblock \doi{10.1109/ISDA.2005.33}.

\bibitem[Zhang et~al.(2021)Zhang, Zhao, and Yao]{ZHANG2021}
X.~Zhang, Y.~Zhao, and X.~Yao.
\newblock Forecasting corporate default risk in china.
\newblock \emph{International Journal of Forecasting}, 2021.
\newblock ISSN 0169-2070.
\newblock \doi{https://doi.org/10.1016/j.ijforecast.2021.04.009}.

\end{thebibliography}


\newpage 
\section*{Appendix} 
\appendix 
\renewcommand{\thesubsection}{\Alph{subsection}}
\subsection{Simple example of CBR}\label{Instance}
Assume there are two persons ($Q$ and $C$) who have three ($L=3$) equally-weighted features ($w_j = 1/3$): weight, height and age as shown in Table \ref{t:instance}. In addition, among all the people (dataset), the highest and lowest value of weight, height and age are known as well. The parameters $a_1$, $a_2$, $a_3$, $b_1$, $b_2$ and $b_3$ in Equation (1) are assumed to be 3, 2, 1, 1, 2 and 3, respectively. The similarity calculation of $Q$ and $C$ based on those features based on Equation (1) and (2) is given: 

\begin{align*}
sim_1(q_1,c_1)= \big(\frac{D_{1}-(q_1-c_1)}{D_{1}}\big)^{\bm{b_1}} = \big(\frac{100-(120-100)}{100}\big)^{\bm{1}} = 0.8\\
sim_2(q_2,c_2)= \big(\frac{D_{2}-(q_2-c_2)}{D_{2}}\big)^{\bm{b_2}} = \big(\frac{50-(180-170)}{50}\big)^{\bm{2}} = 0.64\\
sim_3(q_3,c_3)= \big(\frac{D_{3}-(c_3-q_3)}{D_{3}}\big)^{\bm{a_3}} = \big(\frac{40-(40-35)}{40}\big)^{\bm{1}} = 0.875
\end{align*}
  
\begin{align*}
    Sim(Q,C) = \sqrt{\sum_{j=1}^3 \bm{w_j}\times(sim_j(q_j,c_j))^2} =
\sqrt{\bm{\frac{1}{3}}\times(0.8)^2 + \bm{\frac{1}{3}}\times(0.64)^2 + \bm{\frac{1}{3}}\times(0.875)^2} = 0.7779
\end{align*}

\begin{table}[ht!]
\caption{The features of Person Q and C.}\label{t:instance}
 \begin{threeparttable}
 \begin{tabular*}{\textwidth}{l  p{3cm} p{3cm} p{3cm}  } 
\toprule
&Weight&Height&Age\\
\midrule
Person $Q$&120 ($q_1$)&180 ($q_2$)&35 ($q_3$)\\
Person $C$&100 ($c_1$)&170 ($c_2$)&40 ($c_3$) \\
Maximum &180&200&60 \\
Minimum&80&150&20\\
Difference between Max and Min ($D_j$)&100 ($D_1$)&50 ($D_2$)&40 ($D_3$) \\
\bottomrule
\end{tabular*}

 \begin{tablenotes}
      \item  
    \end{tablenotes}
  \end{threeparttable}
\end{table}

\subsection{PSO}\label{PSO}
PSO is a computational method that optimizes a problem by iteratively improving a solution measured in a certain metric. The basic idea is that a population of particles moves through the search space. Each particle has knowledge about its current velocity, its own past best configuration ($\overrightarrow{p}(t)$), and the current global best solution ($\overrightarrow{g}(t)$). Based on this information, each particle's velocity is updated such that it moves closer to the global best and its past best solution at the same time. The velocity update is performed according to the following equation:

\begin{equation}
\begin{split}
\overrightarrow{v}(t+1) & =\omega \overrightarrow{v}(t) + c_{1}r_1(\overrightarrow{p}(t)-\overrightarrow{x}(t))+  c_{2}r_2(\overrightarrow{g}(t)-\overrightarrow{x}(t)) 
\end{split}
\end{equation}
\noindent where $c_1$ and  $c_2$ are constants defined beforehand, that determine the significance of $\overrightarrow{p}(t)$ and $\overrightarrow{g}(t)$. $\overrightarrow{v}(t)$ is the velocity of the particle, $\overrightarrow{x}(t)$ is the current particle position, $r_1$ and $r_2$ are random numbers from the interval [0,1], and $\omega$ is a constant ($0 \leq \omega \le 1$).
The new position is calculated by summing the previous position and the new velocity as follows:
\begin{equation}
\overrightarrow{x}(t+1) = \overrightarrow{x}(t) +\overrightarrow{v}(t+1)
\end{equation}
In each iteration, if the best individual solution is better than the global best solution, which will be updated by the best individual solution. This iterative process is repeated until a stopping criterion is satisfied. In the proposed CBR system, PSO is used to search for the optimal parameters for each feature similarity function.
\subsection{Parallel computing}\label{Parallel_computing}
Parallel computing is a type of computation where large calculations can be divided into smaller ones, and their computing processes are carried out simultaneously. The potential speedup of an algorithm on a parallel computing framework is given by Amdahl's law, which can be expressed mathematically as follows:
\begin{equation}
    Speedup = \frac{1}{(1-p)+\frac{p}{s}}
\end{equation}
\noindent where $Speedup$ is the theoretical maximum speedup of the execution of the whole task, and $p$ is the proportion of a system or program that can be made parallel and $s$ stands for the number of processors. 

One successful application of GPU-based parallel computing is deep learning, which is a typical intensive computing and training task that can be split. For the CBR querying process, it also can be paralleled. In particular, the similarity calculation between a query and each case can be processed simultaneously. The algorithm for predicting N queries with L features (query matrix) based on M cases with L features (reference matrix) is shown as follows:  \\ 

\begin{algorithm}[H]
\DontPrintSemicolon
  
\KwInput{N$\times$L query matrix, M$\times$L reference matrix, $\bm{a_j}$, $\bm{b_j}$, $\bm{w_j}$, $D_j$ and $\bm{k}$}
\KwOutput{N prediction vector $Prediction_{n}$}
\tcp{Each thread simultaneously calculates each similarity $sim_{n,j}$ between $q_{n,j}$ and $c_{m,j}$, where $n$ = 1, ..., N and $m$ = 1, ..., M.}
\While{calculate the similarity between $q_{n,j}$ and $c_{m,j}$}
  {
  \For{$j:= 1$ \KwTo L}
  {\eIf{$q_{n,j}$ > $c_{m,j}$}
    {
        $sim_{n,j} = \big(\frac{D_{j}-(c_{m,j}-q_{n,j})}{D_{j}}\big)^{\bm{a_j}}$     
    }
    {
    	$sim_{n,j} = \big(\frac{D_{j}-(q_{n,j}-c_{m,j})}{D_{j}}\big)^{\bm{b_j}}$ 
    }
    }
    Synthread() \tcp{Wait for the computing completion for all the similarities $sim_{n,*}$.}
    $Sim_{n} = \sqrt{\sum_{j=1}^L w_j \times sim_{n,j}^2}$
    }
   \tcp{Wait for  the computing completion for all the similarities $Sim_{n}$ for $n_{th}$ query $q_{n,*}$}
   synchronized for query $q_{n,*}$.\;
   Sort and select the $\bm{k}$ most similar cases with $q_{n,*}$ from $Sim_n,\ for\  n = 1, ..., N.$\;
   Voting $\bm{k}$ most similar cases to obtain the prediction for $q_{n,*}$: $Prediction_{n}$.\;
    
\caption{Similarity calculation pseudo code}
\end{algorithm}
\vspace{2mm}
The computation time for querying the test set of the five financial datasets is shown in Table \ref{t:time}. The graphics processing unit (GPU) used in this study is the NVIDIA Geforce GTX 1080. From Table \ref{t:time}, we can observe that the query time increases with increasing magnitude of data. 

\begin{table*}[ht!]
\caption{ Query time instances of the datasets used in cross-validation.}\label{t:time}
 \begin{threeparttable}
 \begin{tabular*}{\textwidth}{p{3.5cm} P{2cm} P{2cm} P{2cm} P{2cm} P{2cm} P{2cm} } 
\toprule
&CCF&CCD&GCA&BC&FD\\
\midrule
Query time (s)      &0.0147 &2.3931 &0.0053 &0.1586 &0.0036 \\
Total cases     &984    &13,272 &600    &4,074  &272   \\
Reference cases &787    &10,617 &480    &3,259  &217    \\
Query cases     &197    &2,655  &120    &815    &55    \\
Features number &30     &24     &20     &10     &83     \\
\bottomrule
\end{tabular*}
 \begin{tablenotes}
      
      \item \scriptsize{Notes: The randomly under-sampling cases data is apportioned into reference and query, with an 80-20 split. The 5-fold cross-validation is used to evaluate the CBR system performance when training model.}
    \end{tablenotes}
  \end{threeparttable}
\end{table*}
\newpage
\subsection{Benchmark models}\label{Benchmark_models}
The benchmark models are briefly introduced as follows:\\

\noindent\underline{Logistic regression:}\\
Logistic regression is a mathematical modeling approach that can be used to describe the relationship of several variables to a dichotomous dependent variable \cite{Kleinbaum1994}. It is an efficient and powerful way to analyze the effect of a group of independent variables on a binary outcome \cite{Stoltzfus20111099}. In logistic regression, regularization is used to reduce generalization error and preventing the algorithm from overfitting in feature rich dataset. The Ridge and Lasso methods are most common used. Consequently, the inverse of regularization strength is also needed to determine. The smaller values specify stronger regularization. The best model can be found by cross-validation grid search. \\

\noindent\underline{K-nearest neighbor:}\\
k-nearest neighbors algorithm is a non-parametric classification method, which means it does not make any assumption on underlying data. It only considers the k nearest neighbors to classify the query point \cite{Mitchell1997}. The hyperparameter required to decide is the k. The best model is achieved through a cross-validation procedure by using a grid search for the k.\\

\noindent\underline{Decision tree:}\\
A decision tree is a map of the possible outcomes of a series of related choices, where each internal choice (node) denotes a test on an attribute, each branch represents an outcome of the test, and each leaf node holds a class label \cite{Song2015}. There are several hyperparameters required to tune. Impurity is used to determine how decision tree nodes are split. Information gain and Gini Impurity are commonly used. The maximum depth of the tree and the minimum number of samples required to be at a leaf node are also important to tune. Cross-validation grid search is applied to find the optimal model.\\
 
\noindent\underline{Gaussian Naive Bayes:}\\
Naive Bayes Classifiers are based on the Bayesian rule and probability theorems and has a strong assumption that predictors should be independent of each other \cite{Mitchell1997}. Gaussian naive Bayes classification is an extension of naive Bayes method with an assumption that the continuous values associated with each class are distributed according to a Gaussian distribution. No hyperparameter tuning is required in Gaussian Naive Bayes. \\

\noindent\underline{Multi-layer perceptron:}\\
A Multi-layer preceptron (MLP) is a class of feedforward artificial neural network (ANN), which consists of at least three layers of nodes: an input layer, a hidden layer and an output layer \cite{GARDNER19982627}. It is a supervised non-linear learning algorithm for either classification or regression. MLP requires tuning a number of hyperparameters such as the number of hidden neurons, layers, and iterations. Hyperband algorithm is used for hyperparameters optimization \cite{JMLR:v18:16-558}.\\

\noindent\underline{Lasso regression:}\\
Lasso regression is a linear regression method that perform both feature selection and regularization in order to enhance the prediction accuracy. The goal of the algorithm is to minimize: $\sum_{j=1}^m(y_j-\sum_{i=1}^nx_{ji}\beta_i)^2 + \lambda \sum_{i=1}^n|w_i|$, where $w$ is the vector of model coefficients and $\lambda$ is a hyperparameter \cite{Robert1996267}. The algorithm has the advantage that it shrinks some of the less critical coefficients of features to zero and $\lambda$ is basically the amount of shrinkage. The best model is selected by cross-validation. 

\newpage
\subsection{Measure metrics}\label{Measure_metrics}
TP (true positive) is the number of correctly classified positive instances. TN (true negative) is the number of correctly classified negative instances. FP (false positive) is the number of positive instances misclassified. FN (false negative) is the number of negative instances misclassified.
\begin{enumerate}[label=(\arabic*)]

\item Precision is referred to as the positive predictive value.
\begin{equation}
    Precision = \frac{TP}{TP+FP}
\end{equation}
\item Recall or sensitivity is referred to as the true positive rate.
\begin{equation}
    Recall = \frac{TP}{TP+FN}
\end{equation}
\item Specificity is referred to as the true negative rate.
\begin{equation}
    Specificity = \frac{TN}{FP+TN}
\end{equation}
\item F1-score or F-measure is the harmonic mean of precision and recall.
\begin{equation}
    F1\mbox{-}score  = 2 \times \frac{precision \times recall}{precision+recall}
\end{equation}
\item ROC\_AUC (the area under the receiver operating characteristic) shows how much a model is capable of distinguishing between classes. Higher the AUC, better the model is. 
\item G-mean is the geometric mean of recall and precision.
\begin{equation}
    G\mbox{-}mean = \sqrt{recall \times precision}
\end{equation}
\end{enumerate}

\newpage
\subsection{Technique for order preference by similarity to ideal solution (TOPSIS)}\label{TOPSIS}

The procedure of TOPSIS can be summarised as follows:
\begin{steps}
\item Calculate the normalised decision matrix. The normalised value $r_{ij}$ is calculated as:
\begin{equation}
      r_{ij} = x_{ij} \bigg/ \sqrt{\sum_{i=1}^{n} x_{ij}^2},\quad i = 1,...,n;\quad j=1,...,m.
\end{equation}
\noindent where $n$ and $m$ denote the number of alternative models and the number of criteria, respectively. For alternative model $A_i$, the performance score of the $jth$ criterion $C_j$ is represented by $x_{ij}$.

\item The weighted normalised decision matrix is calculated as follows:
\begin{equation}
      v_{ij} = w_jr_{ij},\quad i = 1,...,n;\quad j=1,...,m.
\end{equation}
\noindent where $w_{j}$ is the weight of the $jth$ criterion obtained by the information entropy approach. To minimise the input of decision maker, we consider the criterion are equally important.

\item The ideal alternative solution $A^+$ is calculated as:
\begin{equation}
      A^+ = \{v_1^+,...,v_m^+\} = \{(\max_i v_{ij}|j) \in I^{'}),(\min_i v_{ij}|j) \in I^{''} \}
\end{equation}
\noindent where $I^{'}$ denotes benefit criteria and $I^{''}$ denotes cost criteria.

\item The anti-ideal alternative solution $A^-$ is calculated as:
    \begin{equation}
      A^- = \{v_1^-,...,v_m^-\} = \{(\min_i v_{ij}|j) \in I^{'}),(\max_i v_{ij}|j) \in I^{''} \}
\end{equation}
\item The distance of each alternative from the ideal alternative solution $A^+$ is calculated as:
\begin{equation}
      D_i^+ = \sqrt{\sum_{j=1}^m(v_{ij}-v_{j}^+)^2}, \quad i=1,...,n
\end{equation}
The distance of each alternative from the anti-ideal alternative solution $A^-$ is calculated as:
\begin{equation}
      D_i^- = \sqrt{\sum_{j=1}^m(v_{ij}-v_{j}^-)^2}, \quad i=1,...,n
\end{equation}   
\item The relative model degree is calculated as:  
\begin{equation}
      R^+ = D_i^-/(D_i^- + D_i^+), \quad i=1,...,n
\end{equation}   
\noindent where the degree $R^+$ presents the relative rank of alternative models (the larger, the better).
\end{steps}
\newpage
\subsection{The description of the south German credit data}\label{german_data}
The detailed explanation of the features of the south German credit dataset is presented in Table \ref{t:german_description}. 
\begin{table*}[ht!]
\captionsetup{font=scriptsize}
\caption{The description of the south German credit data \cite{Ulrike2019}.}\label{t:german_description}
 \begin{threeparttable}
 \begin{tabular*}{\textwidth}{>{\scriptsize}p{3cm}>{\scriptsize}p{7cm}>{\scriptsize}p{5.6cm} } 
\toprule
Feature&Description&Level\\
\midrule
Status& Status of the debtor's checking account with the bank&1: no checking account, 2: "... < 0 DM", 3: "0<= ... < 200 DM", 4: "... >= 200 DM / salary for at least 1 year"\\
Duration& Credit duration in months &Numeric\\
Credit history& History of compliance with previous or concurrent credit contracts&0: "delay in paying off in the past",
  1: "critical account/other credits elsewhere",
  2: "no credits taken/all credits paid back duly",
  3: "existing credits paid back duly till now",
  4: "all credits at this bank paid back duly"\\
Purpose&Purpose for which the credit is needed&0: "others", 1: "car (new)", 2: "car (used)", 3: "furniture/equipment",  4: "radio/television", 5: "domestic appliances", 6: "repairs", 7: "education", 8: "vacation", 9: "retraining", 10: "business"\\
Amount        &Credit amount in DM &Numeric\\
Savings            &Debtor's savings &1: "unknown/no savings account", 2: "... <  100 DM", 3: "100 <= ... <  500 DM", 4: "500 <= ... < 1000 DM", 5: "... >= 1000 DM"\\
Employment duration&Duration of debtor's employment with current employer &1: "unemployed", 
                      2: "< 1 yr", 
                      3: "1 <= ... < 4 yrs",
                      4: "4 <= ... < 7 yrs", 
                      5: ">= 7 yrs"\\
Installment rate   &Credit installments as a percentage of debtor's disposable income&1: ">= 35", 
                                  2: "25 <= ... < 35",
                                  3: "20 <= ... < 25", 
                                  4: "< 20"\\
Personal status    &Combined information on sex and marital status &1: "male : divorced/separated",
  2: "female 1: non-single or male : single",
  3: "male : married/widowed",
  4: "female : single"\\
Other debtors      & Is there another debtor or a guarantor for the credit? &1: "none",
  2: "co-applicant",
  3: "guarantor"\\

Present residence      & Length of time (in years) the debtor lives in the present residence &1: "< 1 yr", 
                                  2: "1 <= ... < 4 yrs", 
                                  3: "4 <= ... < 7 yrs", 
                                  4: ">= 7 yrs" \\
Property               &The debtor's most valuable property 
&1; "unknown / no property",
  2: "car or other",
  3: "building soc. savings agr./life insurance", 
  4: "real estate"\\
Age                    &Age in years &Numeric\\
Other installment plans&installment plans from providers other than the credit-giving bank &1: "bank",
  2: "stores",
  3: "none"\\
Housing                &Type of housing the debtor lives in &1: "for free", 2: "rent", 3: "own"\\

Number credits  & Number of credits including the current one the debtor has (or had) at this bank &1: "1", 2; "2-3", 3: "4-5", 4: ">= 6"\\
Job             &Quality of debtor's job &1: "unemployed/unskilled - non-resident",
  2: "unskilled - resident",
  3: "skilled employee/official",
  4: "manager/self-empl./highly qualif. employee"\\
People liable   &Number of persons who financially depend on the debtor &1: "3 or more", 2: "0 to 2"\\
Telephone   &    Is there a telephone landline registered on the debtor's name? &1: "no", 2: "yes (under customer name)"\\
Foreign worker  &Is the debtor a foreign worker? &1: "yes", 2: "no"\\

\bottomrule
\end{tabular*}
\end{threeparttable}
\end{table*}


\end{document}